\documentclass[12pt]{article}

\usepackage{times}
\usepackage{natbib}
\usepackage{bibentry}
\usepackage{newunicodechar,graphicx}

\DeclareRobustCommand{\okina}{%
  \raisebox{\dimexpr\fontcharht\font`A-\height}{%
    \scalebox{0.8}{`}%
  }%
}
\newunicodechar{’}{\okina}

\usepackage{amssymb, amsmath}
\usepackage{rotating}
\usepackage{hyperref}
\usepackage{multirow}
\usepackage{array}
\usepackage[dvipsnames]{xcolor}
\usepackage{xr-hyper}
\usepackage{hyperref}
\usepackage[multiple]{footmisc}
\newcommand*{\rowstyle}[1]{
  \gdef\@rowstyle{#1}%
  \@rowstyle\ignorespaces%
}

\newcolumntype{=}{
  >{\gdef\@rowstyle{}}%
}

\newcolumntype{+}{
  >{\@rowstyle}%
}

\title{Using Temperature Sensitivity to Estimate Shiftable Electricity Demand\\ {\Large Implications for power system investments and climate change} }

\author {Michael J. Roberts,$^{1,2,3\ast}$ Sisi Zhang,$^{1}$ Eleanor Yuan$^{1}$, James Jones$^{5}$, \\ Matthias Fripp$^{2,4}$\\
\\
\normalsize{$^{1}$Department of Economics, University of Hawai’i at M\=anoa}\\
\normalsize{2424 Maile Way, Saunders 542, Honolulu, HI 96822, USA}\\
\\
\normalsize{$^{2}$ University of Hawai’i Economic Research Organization (UHERO)} \\
\\
\normalsize{$^{3}$ University of Hawai’i Sea Grant College Program} \\
\\
\normalsize{$^{4}$ Department of Electrical Engineering, University of Hawai’i } \\
\\
\normalsize{$^{5}$ Northern Virginia Electric Cooperative} \\
\\
\normalsize{$^\ast$Correspondence: mjrobert@hawaii.edu.}
}

\date{}

\begin{document} 


\baselineskip19pt


\maketitle 


\thispagestyle{empty}

\newpage

\section*{Summary}
\setcounter{page}{1}

\textbf{Growth of intermittent renewable energy and climate change make it increasingly difficult to manage electricity demand variability. Centralized storage can help but is costly. An alternative is to shift demand. Cooling and heating demands are substantial and can be economically shifted using thermal storage. To estimate what thermal storage, employed at scale, might do to reshape electricity loads, we pair fine-scale weather data with hourly electricity use to estimate the share of temperature-sensitive demand across 31 regions that span the continental United States. We then show how much variability can be reduced by shifting temperature-sensitive loads, with and without improved transmission between regions. We find that approximately three quarters of within-day, within-region demand variability can be eliminated by shifting just half of temperature-sensitive demand. The variability-reducing benefits of shifting temperature-sensitive demand complement those gained from improved interregional transmission, and greatly mitigate the challenge of serving higher peaks under climate change.}

\section*{Introduction}
The cost of conventional electric power systems depends on how much demand fluctuates over time. Installed capacity must exceed peak demand, while base load (minimum demand) limits the amount of capacity that can be fully utilized. Thus, the greater the spread between peak and base load, the greater the average cost per kilowatt hour. Conventional system design strives to optimally balance capacity cost, operation costs, and capacity utilization given demand variability \citep{luss1982operations,dangl1999investment,fripp2012switch}. Managing variability is becoming increasingly difficult due to the growth of intermittent renewable energy and climate change. On a levelized basis, wind and solar photovolatic (PV) are now the least costly sources of power, but have output that fluctuates with weather and sunlight. As a result, demand net of renewable supply is more variable than demand itself. At the same time, climate change is increasing peak demand for cooling on hot summer days \citep{mcfarland2015impacts,auffhammer2017climate}. Transmission, batteries, pumped-water hydroelectric systems, hydrogen, and other storage methods can help to smooth net demand, but these are costly. 

An alternative to these balancing technologies is to make better use of potentially shiftable demand.%
\footnote{Demand can also be curtailed through inconvenience or discomfort, which customers willingly do under successful demand response programs \citep{jessoe2014knowledge,borenstein2005efficiency,borenstein2005long,Wolak2011}. Our focus pertains to demand-side storage or shifting, which may involve costs on the demand side, but not necessarily a loss of utility or discomfort to customers. Location-specific real-time retail pricing and other mechanisms could be used to harness both kinds of demand reshaping.} %
Some uses of electricity may be flexible in terms of when they draw power from the grid. Smart systems promise to automate and coordinate shiftable demands in order to improve comfort and convenience for customers while reshaping demand to be less costly to serve \citep{darby2012,paterakis2015}. A report, commissioned by the Federal Energy Regulatory Commission, investigated the broader issue of demand response in 2009 \citep{ferc2009}. That report focuses mainly on the share of peak load that might be curtailed or shifted in each state using a bottom-up approach that examined each kind of use from each class of electricity consumer. The report pays less attention to potential effects on base load, which could also have considerable value \citep{deMars2020}. A more recent analysis by the National Renewable Energy Laboratory \citep{mai2020} takes a similar approach and considers implications of flexibility for a range on uses, but assume only a limited degree of flexibility (a maximal one hour shift for HVAC), in conjunction with a single model year (2012). These assumptions were embodied in a recent National Academy of Sciences report on decarbonization pathways \citep{pacala2021}. As we look forward toward planning a future grid with more intermittent renewables, what is needed is a broad assessment of the real-time shiftable demand in each location, time of day, and weather circumstance. Such an assessment can be used in conjunction with synchronized weather data to aid assessment of the best locations to site renewables, transmission, and storage investments. Flexible demand could ultimately reduce the costs of decarbonization by reducing storage and transmission costs, and might therefore involve very different investment strategies, even in the near term, as compared to systems that pay less attention to such potential.

How much flexible demand exists? While many different kinds of demand may be shiftable, a significant share of electricity demand is used for cooling and heating, and affordable technologies exist, and can likely be improved, to shift these loads. For example, with sufficient insulation, energy used for air conditioning and space and water heating can be stored in ice, hot water, underground materials, or other forms of thermal storage for hours to days, and in some cases seasonally \citep{arteconi2012,darby2012,hoonyoon2014,hoonyoon2016}. The resulting flexibility of demand timing can be used to flatten demand profiles, or in more complex renewable systems, shift demand toward variable supply. While other flexible uses of electricity exist, heating and cooling demand is substantial and sensitive to ambient temperature fluctuations, and thus estimable over time and space. 

In this paper we estimate potentially shiftable heating and cooling demand using a ``top down" approach that links hourly electricity demand across the continental United States to fine-grained estimates of hourly temperature over space and time. Our approach bears some resemblance to~\citep{denholm2012using} who identify cooling loads by comparing realized demand in warm months to minimum demand on otherwise similar days. Our approach is more explicit and less susceptible to potentially confounding factors that could be associated with low-demand days and hours; it also accounts for heating-related demand. We use a flexible functional form to account for shifts in demand that arise from seasonal, day of week, and time-of-day effects, as well as geospatial variations in climate, which can factor into temperature sensitivity. Conditional on these factors, weather variation is arguably conditionally exogenous---as if randomly assigned---and thereby constitutes a viable natural experiment to identify temperature-sensitive load that might be shiftable. We cross-validate the model by predicting demand in out-of-sample years, and show these predictions to be highly accurate. We then use the model to predict the share of temperature-sensitive demand in each hour and region. 

\begin{figure}[h!]
\centering
\includegraphics[trim=0cm 0cm 0cm 0.7cm, clip=true, width=0.8\textwidth]{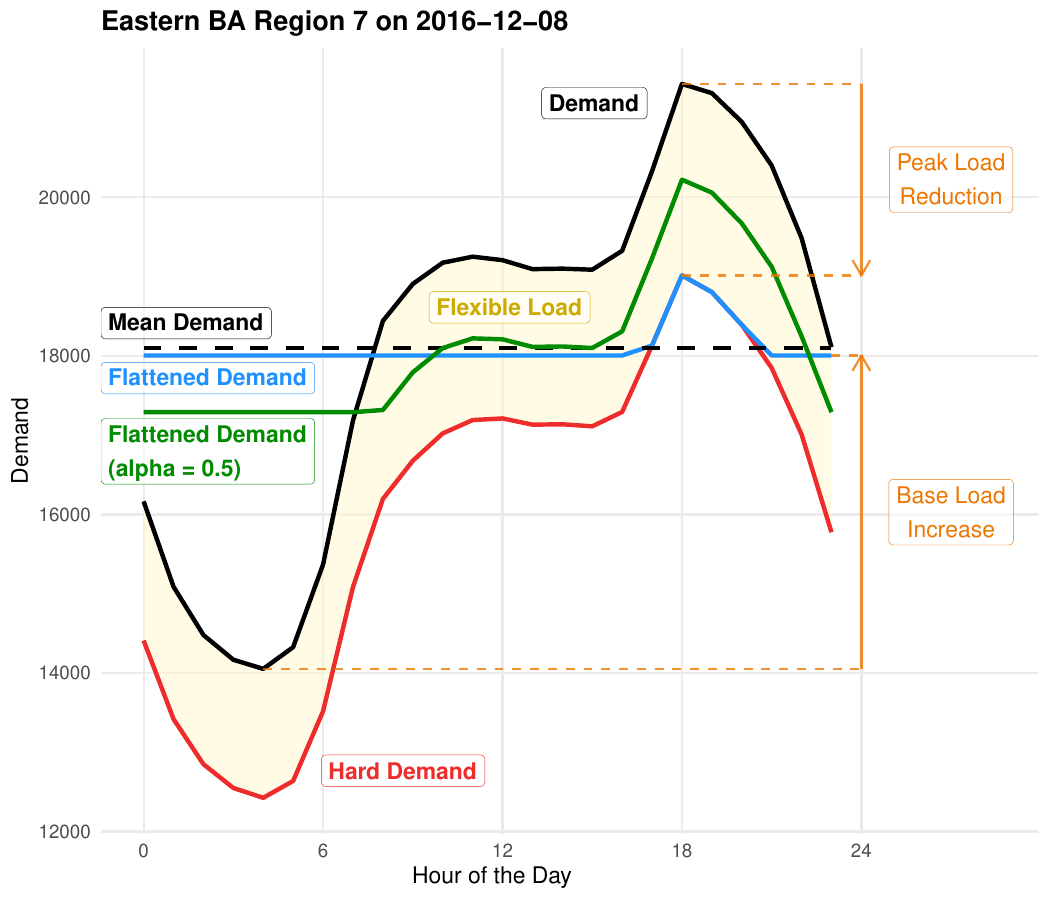}
\caption{\footnotesize \textbf{Demand, Flexible Load, Hard Load, and Flattened Demand.} The graph shows electricity demand for one region in the Eastern Interconnect on December 8, 2016, together with the estimated flexible load (shaded yellow) and remaining hard demand (assumed unshiftable). Flattened demand is constructed by reshaping all (in blue) or half (in green) the flexible load in each hour.}
\label{fig:smoothDemand}
\end{figure}

Assuming different shares ($\alpha$) of the estimated temperature-sensitive load in each hour can be shifted within each day, we determine alternative feasible ``flattened" demand profiles. These calculations are illustrated for one day in one region in the Eastern interconnect, for $\alpha=0.5$ and $\alpha = 1$ (Figure~\ref{fig:smoothDemand}). At least in a conventional system, the degree to which demand can be flattened comports with the cost of the overall system, holding total demand fixed. We show how much reshaping demand in this manner can serve to flatten load with and without transmission across regions within and between the three interconnects in the United States: East, West and ERCOT.\footnote{ERCOT stands for the Electric Reliability Council of Texas.} Finally, we predict demand and estimate shiftable load under a uniform climate change scenario in which all temperatures increase by 2$^{\circ}$C.

\section*{Results}
\begin{figure}[h!]
\centering
\includegraphics[trim=0cm 0cm 0cm 0.7cm, clip=true, width=0.9\textwidth]{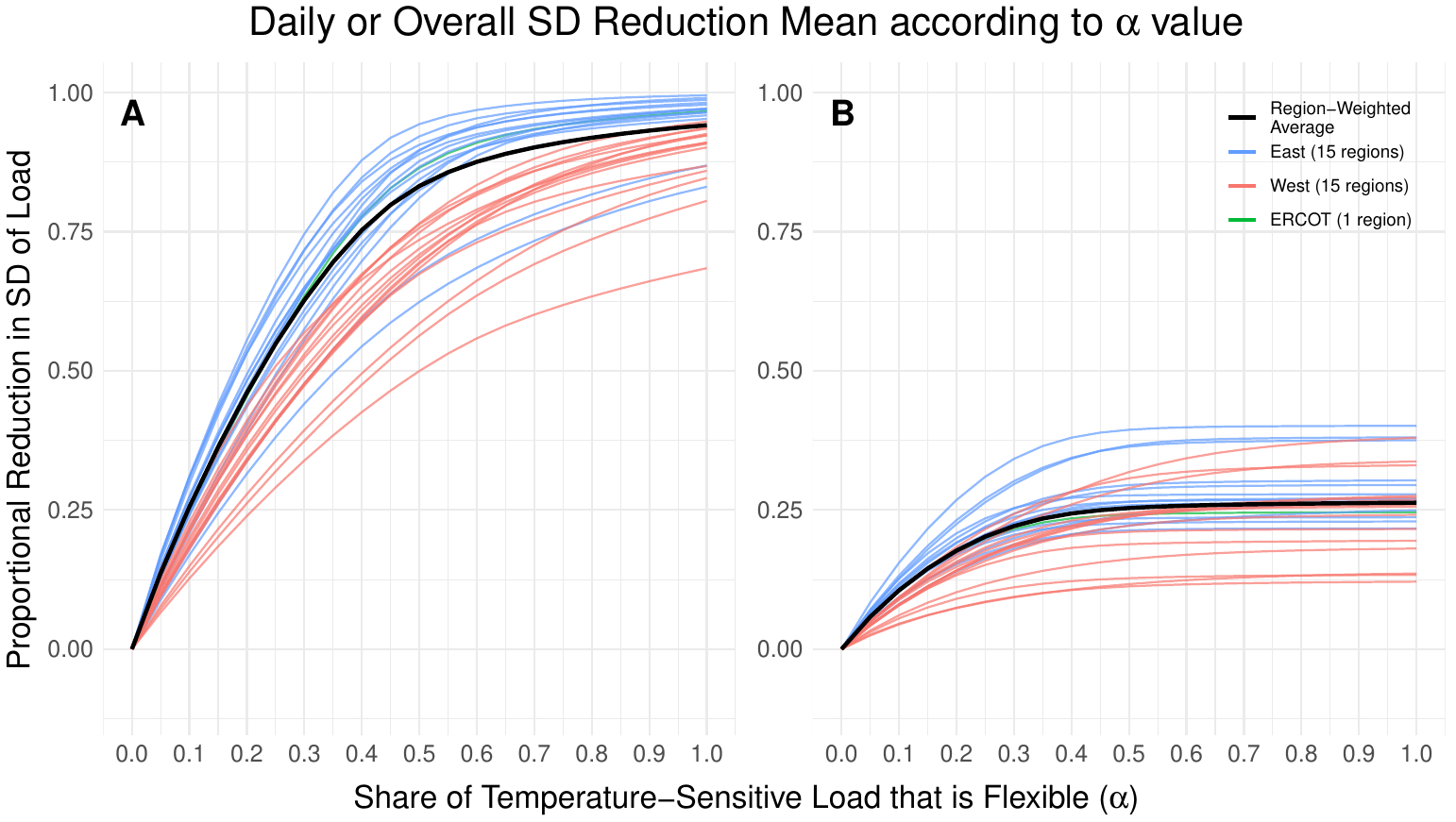}
\caption{\footnotesize \textbf{Proportional reduction in overall load variability for different shares of temperature-sensitive load being flexible.} Each line in the graphs show the reduction in the (A) daily or (B) overall (3-year) standard deviation (SD) of demand for a region when raw demand is optimally flattened using $\alpha$ share of temperature-sensitive load within each day. Each line is colored according to the interconnect in which the corresponding region lies. The thicker black line represents the demand-weighted regional average of daily (A) or overall (B) reduction in SD} 
\label{fig:combo_sd}
\end{figure}

\paragraph{If half the temperature-sensitive load is shiftable ($\alpha = 0.5$), regional daily peak load can be reduced by an average of 10.1\%, daily base load increases by an average of 22.2\%, the daily standard deviation (SD) of load falls by an average of 76.9\%, and 17.9\% of region/days can be flattened completely.}
If all temperature-sensitive load is shiftable in practice, these numbers improve: average daily peak load falls by 13.2\%, average daily base load increases by 24.3\%, the average daily SD falls by 92.2\%, and 60.1\% of region/days can be flattened completely. Thus, roughly three quarters of peak-shaving potential and about 90\% of the base-load increasing potential is achieved if just half the temperature-sensitive load is actually shiftable within days. If only a quarter of temperature-sensitive load is actually shiftable, then roughly half of the potential flatting is possible. All of these statistics are summarized in the top row of Table~\ref{tab:smoothdem_stat}.  To illustrate the full range of what might be possible from demand-side shifting of temperature-sensitive demand, we show the percent decline in the daily SD for the full range of $\alpha \in [0.01, 1]$ (Figure~\ref{fig:combo_sd}). The figure also shows how flatting (SD reduction) varies across regions. See also SD reduction map in Supplemental Information Figure~\ref{fig:sdred_map}. More flattening tends to be possible in Eastern regions as compared to Western regions, while ERCOT (Texas) has somewhat more potential flattening than the national average, typical of many regions in the Eastern interconnect.

\begin{sidewaystable}[h!]
    \centering
    \caption{\textbf{Changes in daily peak and base load and within-day variability when demand is flattened using different levels of $\alpha$ and regional aggregation.}}
     \label{tab:smoothdem_stat}
   \begin{tabular}{l c c c c}
     \hline\hline
     Level of & Peak Load & Base Load & SD & Share of\\
     Connectivity & Reduction & Increase & Reduction & Flattenable Days\\ [0.5ex] 
     \hline 
     {$\alpha$ value} & 0 $\vert$ 0.25 $\vert$ 0.5 $\vert$ 1 & 0 $\vert$ 0.25 $\vert$ 0.5 $\vert$ 1 & 0 $\vert$ 0.25 $\vert$ 0.5 $\vert$ 1 & 0 $\vert$ 0.25 $\vert$ 0.5 $\vert$ 1 \\
     \hline
     \multicolumn{5}{c}{-All Seasons-}\\
     \hline
     Regional BA & 0.0 $\vert$ 5.4 $\vert$ 10.1 $\vert$ 13.2 & 0.0 $\vert$ 16.6 $\vert$ 22.2 $\vert$ 24.3 & 0.0 $\vert$ 48.6 $\vert$ 76.9 $\vert$ 92.2 & 0.0 $\vert$ 0.7 $\vert$ 17.9 $\vert$ 60.1\\ 
     Interconnect & 0.9 $\vert$ 6.4 $\vert$ 11.2 $\vert$ 13.8 & 0.6 $\vert$ 17.7 $\vert$ 23.0 $\vert$ 24.3 & 4.6 $\vert$ 55.7 $\vert$ 84.9 $\vert$ 96.8 & 0.0 $\vert$ 0.4 $\vert$ 24.0 $\vert$ 68.4\\ 
     Nationwide & 2.5 $\vert$ 8.1 $\vert$ 12.7 $\vert$ 14.1 & 2.0 $\vert$ 18.9 $\vert$ 23.3 $\vert$ 23.7 & 12.7 $\vert$ 67.1 $\vert$ 94.3 $\vert$ 99.9 & 0.0 $\vert$ 1.1 $\vert$ 42.5 $\vert$ 95.7\\  
     \hline
     \multicolumn{5}{c}{-Winter—Dec/Jan/Feb-}\\
     \hline
     Regional BA & 0.0 $\vert$ 4.4 $\vert$ 7.8 $\vert$ 10.7 & 0.0 $\vert$ 12.6 $\vert$ 16.1 $\vert$ 17.7 & 0.0 $\vert$ 52.8 $\vert$ 77.2 $\vert$ 92.2 & 0.0 $\vert$ 2.5 $\vert$ 22.1 $\vert$ 59.2\\ 
     Interconnect & 1.1 $\vert$ 4.9 $\vert$ 8.0 $\vert$ 10.9 & 0.8 $\vert$ 13.5 $\vert$ 16.7 $\vert$ 17.8 & 7.7 $\vert$ 58.3 $\vert$ 81.9 $\vert$ 96.0 & 0.0 $\vert$ 1.6 $\vert$ 20.2 $\vert$ 56.9\\ 
     Nationwide & 3.5 $\vert$ 7.7 $\vert$ 10.5 $\vert$ 11.5 & 2.6 $\vert$ 15.3 $\vert$ 17.0 $\vert$ 17.2 & 21.1 $\vert$ 78.1 $\vert$ 95.8 $\vert$ 99.8 & 0.0 $\vert$ 4.5 $\vert$ 54.7 $\vert$ 95.9\\  
     \hline
     \multicolumn{5}{c}{-Summer—Jun/Jul/Aug-}\\
     \hline
      Regional BA & 0.0 $\vert$ 8.1 $\vert$ 15.2 $\vert$ 17.8 & 0.0 $\vert$ 22.8 $\vert$ 30.2 $\vert$ 32.1 & 0.0 $\vert$ 53.5 $\vert$ 86.5 $\vert$ 97.2 & 0.0 $\vert$ 0.0 $\vert$ 30.5 $\vert$ 80.8\\ 
     Interconnect & 0.5 $\vert$ 9.4 $\vert$ 17.1 $\vert$ 18.5 & 0.5 $\vert$ 24.0 $\vert$ 30.6 $\vert$ 31.2 & 1.5 $\vert$ 60.6 $\vert$ 95.1 $\vert$ 99.9 & 0.0 $\vert$ 0.0 $\vert$ 49.7 $\vert$ 97.3\\ 
     Nationwide & 1.5 $\vert$ 10.3 $\vert$ 17.6 $\vert$ 18.0 & 1.8 $\vert$ 24.9 $\vert$ 30.8 $\vert$ 30.9 & 5.9 $\vert$ 65.9 $\vert$ 98.9 $\vert$ 100.0 & 0.0 $\vert$ 0.0 $\vert$ 76.0 $\vert$ 100.0\\ 
     \hline\hline
     \multicolumn{5}{p{7.5in}}{{\footnotesize \emph{Notes}: The table shows the average percent reduction in peak load, average percent increase in base load, average percent reduction in daily standard deviation (SD) of load, and percentage of perfectly flattenable days, when all ($\alpha=1$), half ($\alpha=0.5$), or a quarter ($\alpha =0.25$) of temperature-sensitive load in each hour is shiftable to another hour in the same day.  These calculations are performed for the individual regions, when regions are pooled within each interconnect, and when all regions across the continental United States are pooled (Nationwide). The $\alpha=0$ column shows how much transmission flattens load. The calculations are also broken out for Winter and Summer months.}}
    \end{tabular}
 \end{sidewaystable}

\paragraph{Electricity demand is more strongly associated with cooling degree hours (CDH\protect\footnote{With $CDH = \max \{T - 18, 0\}$, more details see STAR Methods.}) than with heating degree hours (HDH\protect\footnote{With $HDH = \max \{18 - T, 0\}$, more details see STAR Methods.}), while the amount of CDH and HDH varies regionally, with more of both in Eastern regions and ERCOT, which helps to explain why these regions have more temperature-sensitive load and more potential flattenability.}
There are clear seasonal and climatic patterns in temperature-sensitive demand, with more concentrated in Winter and Summer. The summary statistics and regression models show a stronger link between demand and CDH than to HDH, but there tends to be a more HDH on average, while some Southern regions have far more CDH. The greater frequency and response to HDH and CDH in the Eastern interconnect and ERCOT leads to more potential flattening than in the West, which is milder. We show these relationships for ERCOT in Figure~\ref{fig:er_cdhhdh_temp}; other regions and regression results are reported in Supplemental Information, including seasonal patterns. Note that Figure~\ref{fig:er_cdhhdh_temp} does \emph{not} account for seasonality (hour of year), time of day, or day of week, as do the regression results. Table~\ref{tab:smoothdem_stat} summarizes flattenability separately for Winter and Summer months. See also Figure~\ref{fig:seasonal_sd} for seasonal SD reduction in Supplemental Information. Because the overall peak load tends to be in summer, driven mainly by cooling demand, the potential flexibility of this load could be particularly valuable.

\begin{figure}[h]
\centering
\includegraphics[trim=0cm 0cm 0cm 0.7cm, clip=true, width=0.75\textwidth]{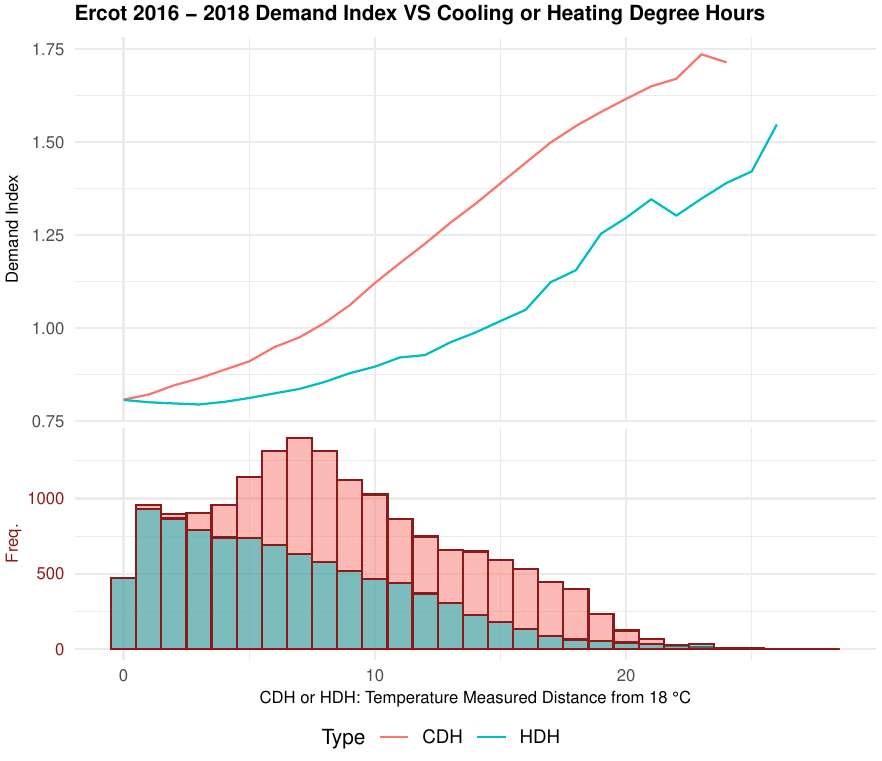}
\caption{\footnotesize\textbf{ERCOT Demand in relation to CDH and HDH.} The graph shows the ratio of electricity demand to mean demand in relation to average CDH and HDH in the hour, each aggregated over all grids in the region. The graph also shows the (overlaid) frequency distributions of CDH and HDH. Texas has more CDH than HDH, plus a stronger association with CDH than HDH. Graphs for the other regions, 15 in each interconnect, are provided in Supplementary Information (Figure S4 and S5).}
\label{fig:er_cdhhdh_temp}
\end{figure}

\vspace{0.5cm}

The measures reported above pertain to individual days and individual regions. There are two key concerns with these measures: (1) electricity can be traded between regions, using existing transmission within interconnects, or in the future, using new transmission within or between interconnects; (2) seasonal and day-of-week variation also matter for system cost, not just variability across hours within each day. To address these issues, we also report potential flattening statistics by interconnect and nationally in Table~\ref{tab:smoothdem_stat}. These measures pool regional demands and potentially shiftable demands in each hour, synchronizing across time zones, and then flattens all regions simultaneously. This allows us to see how much transmission can aid smoothing of demand, with or without demand shifting. We also break out statistics by season and, separately, calculate how much within-day shiftable load can diminish \emph{overall} variability---the SD across the whole three-year (2016-2018) time period examined.

\paragraph{When shiftable load is paired with perfect transmission between regions, average daily load variability (SD) can be reduced by 84.9\% when all regions within each interconnect are pooled and by 94.3\% when all regions are pooled nationwide, assuming just half the temperature-sensitive load is shiftable.}
If all temperature-sensitive load is shiftable, the reductions are 96.8\% and 99.9\%, respectively, and if just a quarter of temperature-sensitive load is shiftable, the reductions are 55.7\% and 67.1\%, respectively. The size of peak load reductions and base load increases associated with an increasing $\alpha$ are slightly lower when load is pooled across regions because pooling demand across regions already reduces the peak and increases base. The share of completely flattenable days increases considerably when load is pooled and then flattened. All of these statistics are reported in Table~\ref{tab:smoothdem_stat}.

\vspace{0.5cm}

We now consider how much within-day shifting of temperature-sensitive load can reduce \emph{overall} variability. Demand varies across days, seasons, and years, not just across hours within each day, and variation over this longer time scale is also pertinent to costs. Temperature-sensitive load is likely only shiftable over a short horizon, and our analysis assumes it is shiftable only within each 24-hour day. This duration of shiftability is greater than what might be achieved through simple thermostat adjustments, but potentially conservative for some kinds of demand-side thermal storage \citep{ruan2016,heine2021}. Nevertheless, we find that within-day flexibility can do a lot to reduce overall variability. 

\paragraph{Assuming half of temperature-sensitive demand can be shifted within days ($\alpha=0.5$), the overall SD of demand for a typical region can be reduced by roughly 25\%.} Larger shares of shiftable temperature-sensitive demand does little to further reduce the overall SD of demand (Figure~\ref{fig:combo_sd} B), indicating that diminishing returns to flexibility arise faster when considering overall variability as opposed to daily variability. We find overall load-flattening potential varies widely across regions, with more generally available in the Eastern interconnect than in the Western interconnect.

\begin{figure}[h!]
\centering
\includegraphics[trim=0cm 0cm 0cm 0.7cm, clip=true, width=0.99\textwidth]{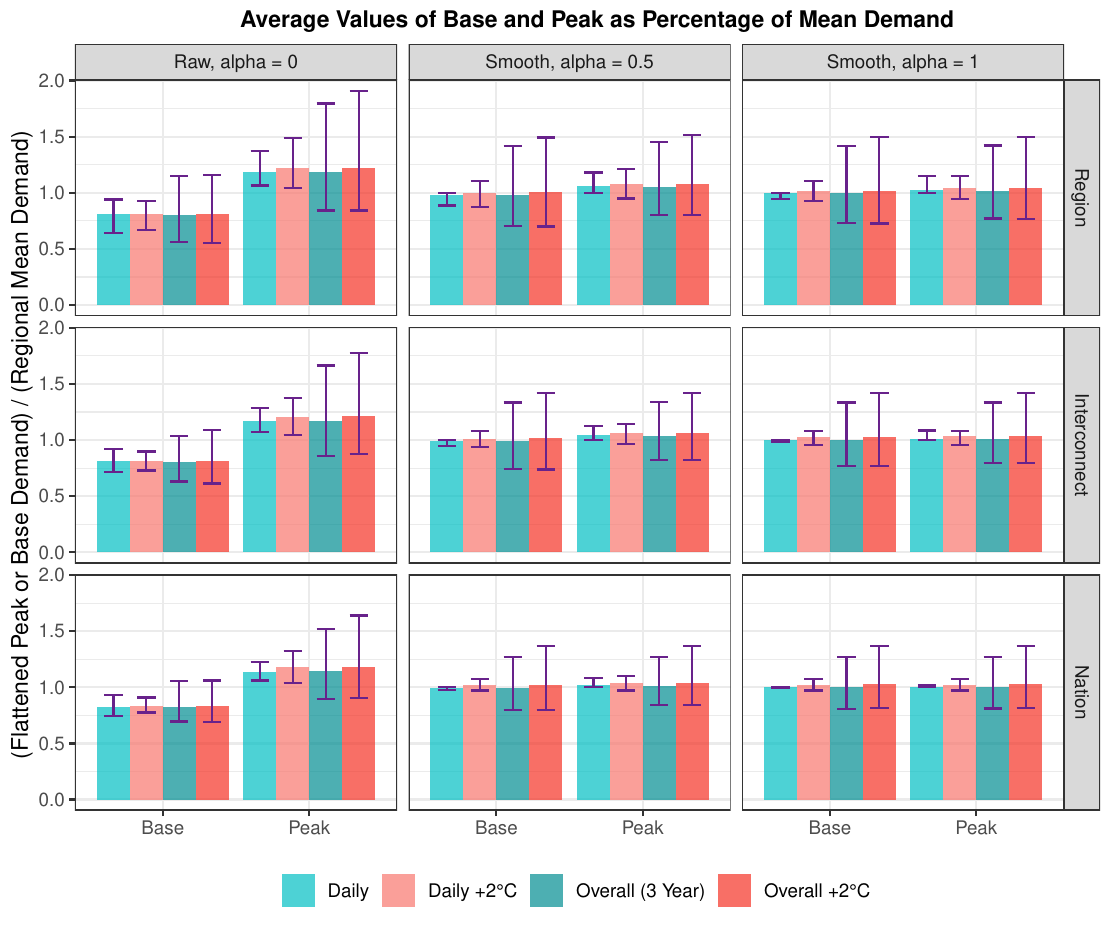}
\caption{\footnotesize \textbf{The influence of demand flexibility, transmission, and climate change on daily and overall base and peak load.} The aqua colored bars show average values of daily peak and base load divided by the same-day mean (lighter shade) or overall (3-year) mean (darker shade). The red bars indicate the same average values with +2$^\circ$C change in temperature, normalized by the actual historical load. Whiskers mark the 1\textsuperscript{st} and 99\textsuperscript{th} percentiles of daily peak and base demand. Demand flexibility increases from left to right, where $\alpha=0$ is raw demand (left column), $\alpha=0.5$ is demand optimally flattened using half the temperature-sensitive load, and $\alpha=1$ is demand optimally flattened using all of the temperature-sensitive load. Transmission increases from top to bottom, where the first row assumes no connectivity between regions, the second row assumes perfect transmission within interconnects (East, West, ERCOT), and the last row assumes perfect transmission across the contiguous United States.}
\label{fig:trans_flex_ALL}
\end{figure}

Because transmission and load shifting both can be used to flatten effective demand profiles, we illustrate the relative importance of each in Figure~\ref{fig:trans_flex_ALL}. This figure is used to illustrate a wide range of results. The aqua-colored bars in the graph show the mean peak and mean base load and the whiskers show the 1\% and 99\% percentiles of peak and base relative to each region's mean load. The more similar peak and base, the flatter the typical load profile. The lighter colored bars show peak and base relative to the \emph{daily} mean, while the darker bars show peak and base relative to the \emph{overall} mean. The choice of baseline mean has little bearing on the size of the bars, but normalizing by the overall mean noticably widens the spread of the whiskers to reflect overall variability across three years not just within-day variability.
The panels of the graph stratify calculations across different $\alpha \in \{0,0.5,1\}$ and three levels of transmission: no transmission between regions, perfect transmission within interconnects, and perfect transmission nationally. Thus, the top-left panel shows the raw data, with no flattening, and the bottom right shows the combined effect of full shiftability of temperature-sensitive demand and perfect transmission across all regions. 

Moving from top-left to bottom right, the bars become nearly equal in height and the whiskers narrow.  While improved transmission does act to reduce the effective peak and increase the effective base (moving from top to bottom panels), we see considerably more flattening arising from shifting of temperature sensitive load (moving from left to right panels). While the \emph{mean} peak and \emph{mean} base loads quickly converge to the overall mean as $\alpha$ increases, indicating strong smoothing overall days, the whiskers on the darker bars indicate that between-day/between-region variability remains, but is markedly diminished. Transmission does more to diminish the spread of the whiskers than equalize the height of the mean peak and mean base bars, both with and without flattening of temperature-sensitive demand. In this sense, improved transmission can complement demand-side smoothing, as we detail below.

\paragraph{Shifting of temperature-sensitive load does more to reduce variability than improved transmission.}
Whereas perfect transmission between all regions can increase the effective mean base load from 80.7\% of the overall mean to about 82.7\% of the overall mean, and reduce the effective mean peak load from 118.2\% of the overall mean to 113.8\% of the overall mean, shifting just half of the temperature-sensitive load ($\alpha=0.5$) increases the effective mean base load to 97.8\% of the overall mean and reduces the effective mean peak load to 105.9\% of the overall mean. With perfect transmission \emph{and} shiftability of all temperature-sensitive load ($\alpha=1$), the effective mean base load can be increased to 100.0\% of the overall mean and the effective mean peak can be reduced to 100.0\% greater than the overall mean; i.e., demand can be perfectly flattened within days. If we look at the 1\textsuperscript{st} percentile of overall base load and the 99\textsuperscript{th} percentile of the overall peak, we find perfect transmission increases the base from 56.4\% to 69.6\% of the overall mean while reducing the peak from 179.4\% to 152.1\% of the overall mean. For comparison, flattening within regions using just half the temperature-sensitive demand increases the base to 70.4\% and reduces the peak to 144.9\% of overall demand. With perfect nationwide transmission and flattening with half the temperature-sensitive demand, the overall base rises to 79.6\% and the overall peak falls to 127.1\%. These statistics and many more are reported in the Supplemental Information (Table~\ref{tab:smoothdem_stat_overall}, Table~\ref{tab:bargraphtable}, Table~\ref{tab:smoothdem_stat_daily_2C} and Table~\ref{tab:smoothdem_stat_overall_2C}), and displayed in Figure~\ref{fig:trans_flex_ALL}.

\vspace{0.5cm}

\paragraph{Shifting of temperature-sensitive demand can mitigate impacts of climate change.}
Our analysis of climate change impacts is straightforward: we shift the historical distribution of temperatures 2$^\circ$C warmer in each hour, and project the change in total demand and share flexible load using the regression model. To illustrate how climate change affects peak load, base load, and overall flattenability of loads over space and time, we include red bars and whiskers in figure~\ref{fig:trans_flex_ALL} that show climate-shifted base and peak loads, normalized by the historical means instead of the climate-shifted means. This normalization allows us to see how mean, base, and peak loads change with climate, both before and after accounting for changing shiftability. The impact of climate change holding all else the same (without shifting), is illustrated in the top-left by the increase in the peak red colored bars relative to the aqua colored bars, and by the increase in the peak top whisker. \textbf{We find a 2$^\circ$C increase in temperature causes average daily peak demand to increase from 118.6\% to 122.5\% of historic daily mean and from 118.2 to 121.9\% of the historic overall mean, while the 99\textsuperscript{th} percentile of peak load increases from 179.4\% to 191.0\% of the the overall historic mean} (top-left panel of figure~\ref{fig:trans_flex_ALL}). These predicted changes are similar to those predicted in an earlier study \citep{auffhammer2017climate}, which estimated a 3.5\% increase in average peak and a 7\% increase in the 95\textsuperscript{th} percentile of peak load for the RCP 4.5 warming scenario. Our estimates imply 3.9\% and 11.0\% increases to the same measures, but our estimates pertain to a slightly warmer climate change scenario than RCP 4.5, owing mainly to the recent baseline of 2016-2018. 

We find that climate change has relatively little influence on base load, increasing from 80.4\% to 81.1\% of the historic daily mean and from 80.7\% to 81.2\% of the historic overall mean. When shifting all flexible load, the climate scenario causes average daily base load to increase from 99.6\% to 102.0\% of the historic daily mean and from 99.5\% to 101.7\% of the historic overall mean.

Climate change also increases the peak of shifted and pooled demands, but by a smaller amount. The average peak of flattened demand increases from 105.5\% / 105.9\% to 107.7\% / 107.9\% of the historic daily/overall mean when $\alpha=0.5$, and from 102.0\% / 102.2\% to 104.2\% / 104.3\% when $\alpha=1.0$. The 99\textsuperscript{th} percentile of the peak demand increases from 118.0\% / 144.9\% to 121.3\% / 151.2\% of the historic daily/overall mean when $\alpha=0.5$, and from 114.7\% / 142.1\% to 114.9\% / 149.4\% of the historic daily/overall mean when $\alpha=1.0$. Thus, reductions in peak load stemming from improving shiftability are far larger than increases from climate change. Table~\ref{tab:bargraphtable} in the supplement reports the 1\%, 99\%, and mean values of base and peak loads using daily and overall mean baselines, with and without climate change, for $\alpha \in \{0,0.5,1\}$, and for all levels of transmission (regional, interconnect, and national aggregation). 

%

\section*{Discussion}
Although there is uncertainty about how much temperature-sensitive electricity load might actually be employed to reshape the timing of demand under appropriate policies, and over what duration of time load might be shifted, plausible use of existing technologies and the timing and volume of temperature-sensitive demand indicate that the potential is large enough to have a profound influence on the cost and resiliency of power systems. This conclusion holds if just a fraction of temperature-sensitive demand is actually shiftable within one day. 

It is likely that the temperature-sensitive demand that we identify derives mainly from use of HVAC systems. The total share of temperature-sensitive load that we estimate (19.0\%) roughly matches the share of demand attributed to HVAC systems by others.%
\footnote{This calculation results from taking the demand-weighted average of $SS_{it}$ across all regions and hours (see Methods).}%
For example, the Energy Information Administration reports that HVAC systems comprise 26\% of use from both residential and commercial sectors while attributing 72\% of total demand to these sectors, implying a total HVAC share of about 18.7\%. Industrial uses, which comprise most of the remaining demand, appear to have little HVAC. Although our estimates cannot breakout temperature-sensitive demand by user type, we can estimate how it varies by location, season and hour of day, which is valuable data that can be used to aid planning the grid of the future. It may be that heating and cooling loads have different shiftability at different costs, such that $\alpha$ or the duration of shiftability may differ by user and building type, as well as the particular technology employed. Our intent here is to show in a transparent way a plausible range of what is likely possible.

A substantial commercial market for demand-side thermal storage has existed for many years and is anticipated to grow substantially with renewable energy.%
\footnote{A report by Allied Market Research estimates a global market for thermal energy storage of \$20.8 billion in 2020, about two thirds of which is demand-side cooling and heating \url{https://www.alliedmarketresearch.com/thermal-energy-storage-market}, and projects it to grow to \$50.3 billion by 2030.}%
A substantial academic literature also continues to flourish.
\footnote{For example, \cite{yan2021thermal} report 448 articles on thermal storage published in \emph{Applied Energy} over just ten years, 2009-2018.}%
If thermal storage can shift energy demand relatively cheaply, it raises the question of why these systems have not been adopted more broadly and incorporated into system planning. The answer likely stems from a combination of transactions and retrofit costs, system benefits, and incentives, which vary widely across jurisdictions. Incentives to store energy on the demand side of power systems (``behind the meter") depend on retail pricing schedules, which are regulated. Retail prices need to vary significantly enough for customers to profit from using storage to buy more when prices are low and less when prices are high. Some utilities offer ``time of use" prices, critical peak pricing, compensation for demand reduction during the system peak or to provide regulating reserves, and in rare cases, location-specific real-time marginal cost pricing (LMP) \citep{albadi2008}. Availability of these tariffs, which are regulated by state Public Utilities Commissions, varies widely across utilities and customer classes, and may or may not reflect the full value of storage to the system. 

While LMP might instill appropriate incentives, it tends be resisted by regulators.%
\footnote{LMP would only instill appropriate incentives for demand-side storage in an idealized, perfectly competitive market, which the electricity sector is not. Given the regulated nature of electricity systems, even in situations where LMP has been implemented, demand-side investors would need to have reasonable expectations about centralized decisions, which, among other departures from perfect competition, could include subsidized storage (e.g., such as batteries paired with renewable energy installations) and regulated capacity decisions. Of course, demand-side storage systems can also receive subsidies.} 
Part of the resistance to LMP likely comes wary customers and regulators who have no knowledge of the potential savings and reticence of state public utility commissions that may fear public backlash from extreme spikes in prices, like those that occasionally arise in Texas, which embraces markets more than most States \citep{departo2019}.  At the same time, current regulatory practice sometimes rewards (implicitly) high-cost, centralized solutions that require more capital expenditure, some of which may be avoided with effective demand-side management. This may help to explain why utilities rarely market LMP pricing to customers, even when such contracts are made available \cite{barbose2004}. Nevertheless, there has been notable long-term success with LMP for commercial customers some markets, like Georgia Power \citep{barbose2004}. Because investments in thermal storage and other devices that could enable demand-side flexibility are long lived, the electricity customers that install them may need assurance about their ongoing value, which involves both technological and regulatory uncertainty, which could create further resistance to adoption of LMP and enabling devices like thermal storage. A recent decision by the Supreme Court, however, paired with a new rule by the Federal Energy Regulatory Commission \href{https://www.ferc.gov/sites/default/files/2021-03/E-1.pdf}{(FERC Order No. 2222)}, may facilitate greater use of demand side resources, despite limited availability of LMP and other resisting factors. These rulings should enable independent integrators to harness demand-side resources directly from customers and sell them into wholesale markets as virtual power plants. As state public utilities commissions begin to implement the new rules, system-wide stakeholders and policymakers will need to anticipate (and might beneficially encourage) these impending changes and adjust broader investment decisions accordingly. The estimates we present here might aid this planning.

Invariably, as this study shows, there is joint dependency between the value of energy storage and other investments in the power system, including the capital cost and flexibility of generation resources, the cost of fuels, transmission, and the cost of centralized storage. While far from a comprehensive assessment of benefits and costs, this study provides some sense of scope and scale of demand-side resources related to heating and cooling demand as well as calibrated data resources that can be used to better assess this potential value in conjunction with other investments. In particular, because future power systems are likely to have much higher shares of low-cost intermittent renewables, shiftability is likely to be more valuable than suggested here, but will require different and far more complex kinds of shifts. Instead of simply flattening the demand profile, demand will need to be shifted toward variable supply. The value of transmission in such a system will also be greater, but different and more complex, because some regions of the country have more renewable energy potential than others, due to heterogeneity in available wind, solar radiation, and land resources. Diversity of renewable resource assets will also be valuable as this will serve to even out supply and thereby substitute for storage and demand flexibility. In evaluating the best mix of investments, inclusive of the demand side, it is important to have a synchronized account of supply (wind and solar), demand, and demand flexibility, all of which are all weather dependent and typically associated.

Thus, the nature of shifts that will be most valuable in the system of the future will be closely intertwined with the nature and locations of investments in renewable energy resources and new transmission builds. As a result, the time and location-specific estimates that we have developed for potentially shiftable demand should be paired with other estimable sources of shiftable load, like electric vehicle and hot water demand, and incorporated into state-of-the-art capacity expansion models of the future electric grid. To our knowledge, few capacity expansion models are well-equipped to handle intermittent renewables, much less incorporate a demand side with varying amounts of shiftable load. One exception is GenX, a state-of-the-art capacity expansion model developed by MIT and Princeton, which has been used to co-optimize power-system investments while assuming demand shiftability as estimated by the National Renewable Energy Laboratory \citep{mai2020}. As we noted above, this study develops ``ground up" estimates of electricity demand for a single year (2012) that may not align with actual realized demand. They also assume that HVAC demand can be shifted only 1 hour, while thermal storage technology can facilitate considerably larger shifts. Another exception is SWITCH, the open-source software developed by \cite{johnston2019switch}, which can also account for intermittent renewables, storage, and the full joint-dependency of supply and demand. \cite{imelda2018variable} use this model in a capacity expansion model to estimate the high potential value from LMP pricing that would employ both supply and demand sides of the electricity market for balancing.\footnote{This study indicates a very high value from shiftable demand, much of which is identified as HVAC load, which is assumed to be shiftable within a day, as we do here.} This particular application is narrow in geographic scope and therefore cannot consider the full range of possibilities on a continental scale where transmission and larger seasonal variation will present different challenges and opportunities. Nevertheless, it is likely that such flexibility will have a powerful influence on least-cost decarbonization plans, including the best timing, sizing, and locations of near-term investments like transmission, renewable energy installations, and various forms of storage.

\section*{Limitations of the Study}

Although we do not consider the costs of thermal storage here, including thermal loss and capital expense, they do appear modest \citep{heine2021}. Cost-effective installation of thermal storage systems would likely be paired with installation of HVAC systems that might also improve efficiency and save costs, like air or ground-sourced heat-pump systems. The incremental cost of adding thermal storage will likely vary widely depending on location, scale, and whether it is a new building, a retrofit of a full HVAC system, or simply an addition of storage to an existing system. Other factors affecting costs include the durability of the system and its maintenance costs. Presumably the costs of such systems and their maintenance would depend on the scale of adoption, potentially enabling economies scale and learning-by-doing.

Other kinds of flexible loads, besides those that we can identify statistically, are likely shiftable using demand-side thermal storage. Water heating might be the easiest current use of electricity that is shiftable \citep{buescher2015smart,passenberg2016optimal,ali2019optimizing}. While some water-heating demand may be captured by our estimates, the timing of water-heating demand tends to be more closely associated with use, not with ambient temperature, and thus less estimable in this manner. The Energy Information Administration reports that water heating comprises 10\% of residential electricity demand and 2\% of commercial demand \citep{eia2014}. Electrification of gas water heaters might grow this share, while adoption of heat-pump water heaters could improve overall efficiency while maintaining and growing shiftability. In time, electric vehicles (EVs) could be an even more flexible source of demand than water heating. EVs can be charged at strategically appropriate times in an automated fashion, not only to shift loads, but to help regulate frequency and even provide power to the grid under certain conditions (``vehicle to grid") \citep{noel2018optimizing,zhang2020}. Some kinds industrial processes might also be shifted at low cost, while some heavy computing loads have already been shifted.\footnote{The widely noted success of Georgia Power's real-time pricing tariffs for commercial customers has been attributed mainly to shifting of industrial loads \citep{barbose2004}. Also see Google's effort to shift server demand in an effort to reduce pollution emissions, \href{https://www.blog.google/inside-google/infrastructure/data-centers-work-harder-sun-shines-wind-blows/}{https://www.blog.google/inside-google/infrastructure/data-centers-work-harder-sun-shines-wind-blows/}.} Other techniques would be required to estimate such loads over time and space. In addition, electrification of heat, which has become more economic with improved electric heat pumps, could make Winter demand much higher, but potentially more flexible, and will likely be a necessary component of future decarbonization efforts. Indeed, ground-source heat pump systems, which are effective even in the coldest US climates, would embody some degree of thermal storage over a seasonal scale (not considered here). Thus, the response to HDH is likely grow, likely in a manner that is estimable over time and space. A natural extension of this study could use the relationships to estimate how and where demand and demand shiftability will increase with further electrification of heating systems. For all of these reasons, even the high-end ($\alpha=1$) estimates of potential flexibility estimated here could be conservative for a forward-looking decarbonization plan.

Our analysis cannot distinguish between different types of electricity users, but given commercial and residential demand comprise similar shares of both total demand and HVAC demand, plus the possibility that some thermal loads derive from industrial demand, it is likely that half or more of temperature-sensitive demand derives from large, non-residential users.\footnote{The Energy Information Administration's Annual Energy Outlook, reports that 37\% of electricity demand is residential, 35\% is commercial, and 27\% industrial, and the rest mainly transportation \citep{eia2014}. Also see \href{https://www.epa.gov/energy/electricity-customers}{https://www.epa.gov/energy/electricity-customers}).} Such users are likely to operate at a scale that would make investment in power-shifting devices more economic and thus more likely to be adopted under appropriate incentive policies, although we do not have data in this study to decompose commercial, industrial, and residential customers, much less size classes.

\section*{Author Contributions}
M.R. and S.Z. conceptualized the study and wrote the manuscript; S.Z. conducted the formal analysis, including cross-validation exercises for model selection; E.Y. developed the visualizations and tables; J.J. developed code to clean, aggregate, and integrate the demand data, population data, and weather data; M.F. aided conceptualization of the study.

\section*{Acknowledgments}

\section*{Declaration of Interests}
The authors declare no competing interests. \\

\subsection*{Method Details}

\paragraph*{Data.}
All data used in this analysis are publicly available. All code used in data compilation and estimation will be made publicly available upon publication. 

Cleaned hourly electricity demand from the Energy Information Administration (EIA) form 930, were obtained from \cite{ruggles2020} for all balancing authorities for three calendar years, 2016-2018. For both the Eastern and Western interconnects, we aggregated hourly load into 15 regions. We aggregated balancing authorities based on spatial overlap of coverage areas as indicated in shape files obtained from the Department of Homeland Security.\footnote{See \href{https://hifld-geoplatform.opendata.arcgis.com/datasets/control-areas}{https://hifld-geoplatform.opendata.arcgis.com/datasets/control-areas}.} We found the overlapping area of each balancing authority with every other balancing authority and merged those with largest area overlapping. This process was repeated until 15 balancing authorities remained in Eastern and Western interconnects. The idea with aggregating balancing authorities in this manner was to better match geospatial variation in weather to reported demand, while preserving areas likely connected via transmission. These aggregated regions are plotted in figures below. Note that the aggregated regions still have areas of overlap with other regions.

Gridded air temperature ($T$) data at two meters of elevation (``air.2m") were obtained from the the National Oceanic and Atmospheric Administrations North American Regional Reanalysis data (NARR).\footnote{See \href{ftp://ftp.cdc.noaa.gov/Datasets/NARR/monolevel/}{ftp://ftp.cdc.noaa.gov/Datasets/NARR/monolevel/}.} These data give three-hour measures on a roughly 30 kilometer grid. We linearly interpolate the three-hour data to convert to hourly data.  Two key weather measures are used, cooling degree hours ($CDH$) and and heating degree hours ($HDH$), where 
  
\begin{equation} 
CDH = \max \{T - 18, 0\}
\end{equation} 
\noindent and
\begin{equation} 
HDH = \max \{18-T, 0\}
\end{equation} 
  
To merge these temperature measures with the regional demand data, we overlaid the gridded weather data with roughly 1 km gridded population data, and calculated population-weighted averages over each region.\footnote{Gridded population data were obtained from  \href{http://sedac.ciesin.columbia.edu/data/set/usgrid-summary-file1-2010}{NASA's Socioeconomic Economic Data and Applications Center}.}  Given the non-linearity of CDH and HDH, we first calculate the measures for each NARR grid cell before aggregating to each region. If we were to instead find the average temperature in a region and then calculate CDH and HDH, estimates would be biased. For example, the mean temperature in a region for a particular hour might be 18$^{\circ}$C, while some areas within the region may be cooler and some areas warmer, giving rise to positive CDH and HDH. If we were to first average temperature and then calculate CDH and HDH, both measures would be spuriously measured as zero.

To illustrate the link between demand and temperature in each region, we plot the average demand anomaly (demand divided by mean demand) in each region at each average temperature, breaking out separate measures for each season (Figures ~\ref{fig:er_season_temp} - \ref{fig:w_season_temp}). The graphs of regions in the Eastern and Western interconnects are ordered such that the warmest regions are in the top left and the coolest region is in the bottom right of the panel of graphs.  This ordering is apparent from the frequency histograms of temperatures plotted below each region's panel. The graphs generally show a ``U" shape of load in relation to temperature, with a minimum near 18$^{\circ}$C, which is consistent with convention. The ``U" shape, with a lightly flatter slope near the minimum, which may result from spatial averaging as suggested above, as well as from temperature associated time-of-day effects. We also show each region's relationship with HDH and CDH in Figures~\ref{fig:e_cdhhdh_temp} - \ref{fig:w_cdhhdh_temp} (ERCOT is shown in the body of the paper), which appear slightly more linear. While the slopes in relation to HDH and CDH are similar across regions, they do differ. This may occur because different regions use different sources of heating and cooling, due to differing shares of industrial or other non-temperature sensitive loads, and because different regions experience  different climates. For example, in typically temperate climates that rarely experience very hot temperatures, fewer customers may choose to install air conditioning, and such customers would respond less to extreme temperatures when they do arise. In the Eastern interconnect, we see a stronger link between HDH and relative demand in warmer areas than in colder areas, presumably because colder areas in the Northeast are less likely to use electric heat. Also, homes in the Northeast are presumably better insulated than homes in the South that experience milder Winters. 

\paragraph*{Regression Model.}
The regression models we use for estimating temperature-sensitive load employ a flexible functional form linking demand to hour of day (hod), hour of year (hoy), day of week (dow), and standard weather metrics, heating degree hours (HDH) and cooling degree hours (CDH). Each region was estimated separately. A simplified representation of the functional form is:

\begin{equation}\label{eq:main-reg}
    d_{t} = s_h(hod_t)*s_d(hoy_t)*I(dow_t) + \alpha_h HDH_t +\alpha_c CDH_t + u_t
\end{equation}

\noindent where $d_t$ is the natural log of hourly demand, $s_h(hod_t)$ is a natural spline of the hour of day (1-24) in period $t$, $s_d(hoy_t)$ is a natural spline of the hour of year (1-8760), and $I(dow_t)$ is a vector of day-of-week indicator variables that take on a unique fixed value for each day of week. The `$*$'s indicates that these explanatory variables are interacted, such that separate $s_h()$ and $s_d()$ splines are, in effect, estimated for each day of the week, and the interaction between $s_h$ and $s_d$ implies that hour-of-day effects are allowed to change smoothly over seasons. That is, each column of a spline design matrix is multiplied with each column of the other spline design matrix, with an additional coefficient estimated for each interaction. %
\footnote{We can formalize the complete model as follows. Define $D^h$ and $D^d$ as the design or basis matrices for $s_h()$ and $s_d()$, which have dimensions $N \times K$ and $N \times L$ when $s_h()$ has $K+1$ knots and $s_d()$ has $L+1$ knots. Denote the $k^{th}$ column of $D^h$ as $D_k^h$, and the $l^{th}$ column of $D^d$ as $D_l^d$. Finally, redefine $I$ as an $N \times 7$ matrix of indicator variables for the day of the week, and $\circ$ as a Hadamard product. The resulting regression model is:  
$$
d_t = \sum_{k \in 1,K}\sum_{l \in 1,L}\sum_{m \in 1,7} \beta_{k,m,l} D_k^h \circ D_l^d \circ I_m + \alpha_h HDH_t + \alpha_c CDH_t + u_t
.$$
}
Model selection mainly comes down to a choice of the number of knots in each spline. These were selected by 3-fold cross validation, wherein the model was estimated three times, with two of the three years used in estimation, and one full year of data reserved to assess out-of-sample fit. We select the number of knots based on out-of-sample fit, and select a separate optimized value for each region. For example, region 1 in the Eastern interconnect, summarized in Table~\ref{tab:reg_one} uses 19 knots for the hour-of-day spline and 6 knots for the hour-of-year spline. The number of degrees of freedom for each spline equals the number of knots. Accounting for the interactions, the total number of parameters estimated equals ($19 \times 6 \times 7 + 2 = 800$), which is relatively parsimonious given we observe over 26,000 observations for each region ($3 \times 365 \times 24)$

The summary statistics presented above suggest that climate may influence sensitivity of demand to weather \emph{across} regions, and these differences will be accounted for since we estimate separate models for each region.  We also consider a specification that estimates heterogeneous responses to weather \emph{within} each region. For each 30km weather grid cell we multiply the cell's 12-year mean CDH by realized hourly CDH, and similarly calculate each cell's mean HDH and multiply by hourly HDH. We then aggregate these interactions to the region level by taking the population-weighted average, as we did with CDH and HDH. These covariates allow the marginal responses to CDH and HDH to change linearly in relation to climate measured as the mean of CDH and HDH. As with CDH and HDH, it is important to account for this interaction on the smallest geographic temperature measure before aggregating. 

The regressions generally show that a large share of the residual variance remaining after flexible spline controls for hour-of-year and time-of-day was explained by CDH and HDH. These two weather variables alone typically reduce the
RMSE by about 50\%, and increase the out-of-sample R$^2$ from about 0.6 to 0.9. Using Newey-West robust standard errors to account for the autocorrelated error, find t-statistics on the order of 250 for these weather variables. When we add the interaction terms to the model, fit tends to improve slightly, increasing the out-of-sample R$^2$ by 0.01 or less, but the size and sign of the interaction variables differs greatly across regions. In some cases, for some sub-regions, these interaction terms cause the marginal effect of CDH or HDH to turn negative. We believe some of these results are likely spurious, stemming from the coarseness of our weather measures in relation to population and the rough link between population and demand. Nevertheless, we found inclusion of the CDH and HDH in the regressions have little influence on the estimated share of flexible demand. The interactions did, however, cause some highly implausible projections in a couple regions under climate change, so we opted to use the simpler model without these interactions for the main results. Table~\ref{tab:reg_one} summarizes regression results for one region and Figure~\ref{fig:reg_all_regions} summarizes the regressions for all regions.

\paragraph*{The Share of Temperature Sensitive Demand.}%

Seasonal (hour-of-year), day-of-week, and hour-of-day patterns in electricity demand account for 60\% of demand variance ($r \approx 0.75$) over time for a typical region, but this does vary across regions. This measure is derived from the out-of-sample fit (R$^2$) from a \emph{baseline} regression of hourly demand on a hour-of-year spline interacted with an hour-of-day spline and day-of-week indicator variables, as described above. The interactions of time-of-day and day-of-week allows the time-of-day effects to be different for each day of the week, and interactions with hour-of-year allow these effects to change smoothly over the seasons. We select the number of knots in each spline using cross-validation, reserving a full year of out-of-sample data to assess fit in each fold. All reported fit metrics pertain to the out-of-sample predictions in comparison to actual demand.

Remaining demand variability is driven mainly by weather. Although weather likely underpins much of the seasonal and hour-of-day variation in demand, we explicitly control for these variations because they also derive from other factors, and could confound the weather link. Weather anomalies, in contrast, isolate as-if random variations from which we are more likely to infer a causal link. Demand variability related to short-term fluctuations in temperatures are also more likely to be shiftable. To estimate this relationship, we find population-weighted exposure of cooling and heating degree day hours (CDH and HDH), as well as overall average heating and cooling degree day hours in each location to account for climate. The weather measures are derived from 3-hour, 30 km resolution climate reconstruction data, interpolated to hourly measures. Because degree-day measures are nonlinear in temperature, we first calculate HDH and CDH for each grid cell and population-weight the grid-cell measures to derive regional aggregates. To allow sensitivity to change with climate, we interact long-run average measures (12 years) with hourly outcomes at the grid level, and aggregate the interactions. Adding HDH and CDH to the specification improves the out-of-sample fit to over 90\% in most regions ($r \approx 0.95)$. Adding the interactions with climate can improve fit, but modestly, and has no apparent influence on the results. We therefore focus on results from the simpler model without climate interactions. Tables and illustrations of fit and sensitivity to weather and climate variables are reported in Supplemental Information (Figure~\ref{fig:trans_flex_ALL_interact}, Table~\ref{tab:reg_one}, Table~\ref{tab:smoothdem_stat_daily_interact}, Table~\ref{tab:smoothdem_stat_overall_interact}).

We then estimate, for each region, what electricity demand would have been with zero heating or cooling degree hours (i.e., a constant temperature of 18$^{\circ}$C). We call this estimate \emph{hard demand}. If we define $Y_{it}$ as the observed demand in region $i$ and hour $t$, $\widehat{Y_{it}} \vert T=T_{it}$ as the model-predicted demand, and $\widehat{Y_{it}} \vert T=18^{\circ}C$ as the predicted value at a constant 18$^{\circ}$C, then we estimate the maximum possible \emph{share of shiftable load} in that region and hour as

\begin{equation}
    SS_{it} = \alpha \frac{(\widehat{Y_{it}} \vert T=T_{it}) - (\widehat{Y_{it}} \vert T=18^{\circ}C)}{(\widehat{Y_{it}} \vert T=T_{it})},
\end{equation}

\noindent where the parameter $\alpha$ is the share of temperature-sensitive load that is assumed to be actually shiftable. 

\paragraph*{Flattened Demand.}
To determine ``flattened" demand profiles for each region and day, we developed a script that sorts hourly \emph{hard} demand (as described above) from lowest to highest, and then adds flexible load to hard demand, filling and leveling across hours, until all flexible load has been exhausted. Remaining hours---those with the highest hard demand---are not flattenable.\footnote{The computational script simply adds a day's total flexible demand to the cumulative sum of hard demand sorted from lowest to highest, and divides this cumulative sum by integers one through 24. This vector of length 24 will decline until all flexible load has been distributed to the lowest hard-demand hours, such that the value at the minimum equals the flattened minimum. All scripts will be made public.} If peak hard demand is less than mean demand, then the day is completely flattenable---a constant load is achievable. We executed this script for each day in each region.

These flattened demand profiles are not necessarily the most desirable reshaping of demand that would arise from an optimized power system. They are presented as emblematic of the nature and scale of reshaping that is potentially achievable and could significantly reduce costs in the conventional power systems that currently predominate.

For each region and each hour, we calculate the degree to which electricity load can be flattened for $\alpha \in \{0.01, 0.02,... , 1\}$. For each region and day, we calculate the percent reduction in the standard deviation (SD) of load, the percent reduction in peak load, and the percent increase in base load, when demand is optimally flattened using available shiftable load ($SS_{it}$).\footnote{The percent reduction in SD is calculated as $100\% \times (1 - \frac{SD(\text{flattened demand})}{SD(\text{raw demand})})$.} These calculations are made on both daily and overall (years 2016-2018) time scales.

\paragraph*{Effects of Improved Transmission.}

Demand profiles across regions is imperfectly correlated due to differing time zones and weather. Because transmission can be used to spread supply over larger geographical areas, it effectively flattens the relative variation in total demand while taking advantage of scale economies and regions with comparative advantages in generation at any point in time. We therefore consider load profiles at three geographic scales: (1) regional aggregations of balancing authorities (15 regions in each of the Eastern and Western interconnects, and one in ERCOT, described above); (2) interconnect-level aggregation (Eastern, Western, and ERCOT), which assumes perfect transmission within each interconnect; and (3) national-level, which assumes prefect transmission between all three interconnects that comprise the lower-48 states. While power flows at the regional level currently face fewer constraints, transmission between the interconnects is extremely limited. Comparison across these different levels of aggregation provides some insight into the potential value of demand-side flexibility relative to improved transmission within and between interconnects.\footnote{Transmission has benefits beyond variability management. It also allows for transfer of power from resource-rich areas, like sunny deserts, rich in solar power but with few alternative land uses, to resource-poor areas, like cloudy and densely populated urban areas.} 

To find flattened demand curves at interconnect and national levels, we synchronized hard and flexible demand in each hour and aggregated across candidate regions, and then flattened as described above.

\paragraph*{Climate Shifted Demand.}
We estimate raw demand under simple $+2^\circ C$ climate change as 
\begin{equation}
    Y^{+2^\circ C}_{it} = \left( \widehat{Y_{it}} \vert T=T_{it}+2^\circ C \right) + \widehat{u_{it}}
\end{equation}

\noindent where $\widehat{u_{it}}$ is the residual from the regression model in equation~\ref{eq:main-reg}. We calculate the share of shiftable load under climate change in the same manner as historical demand, 

\begin{equation}
    SS^{+2^\circ C}_{it} = \alpha \frac{(\widehat{Y_{it}} \vert T=T_{it}+2^\circ C ) - (\widehat{Y_{it}} \vert T=18^{\circ}C)}{\widehat{Y_{it}} \vert T=T_{it}+2^\circ C }.
\end{equation}

\noindent Given the amount of warming that has already taken place, adding $+2^\circ C$ to historical values implies substantial warming relative to the pre-industrial period.

\newpage

\clearpage
\bibliography{shift}


\clearpage
\section*{Supplemental Information}
\renewcommand{\thefigure}{S1}
\begin{figure}[!h]
\centering
\caption{\footnotesize \textbf{ERCOT demand in relation to temperature for each season, related to STAR Methods.} The average demand index value according to temperature and season is displayed above a frequency count of temperature. Demand index values are obtained by scaling demand according to regional average demand. The dotted blue line shows 18$^{\circ}$C, the benchmark for determining a cooling or heating degree hour. Compiled hourly data are from 2016 through 2018.}
\includegraphics[trim=0cm 0cm 0cm 0.7cm, clip=true, width=0.9\textwidth]{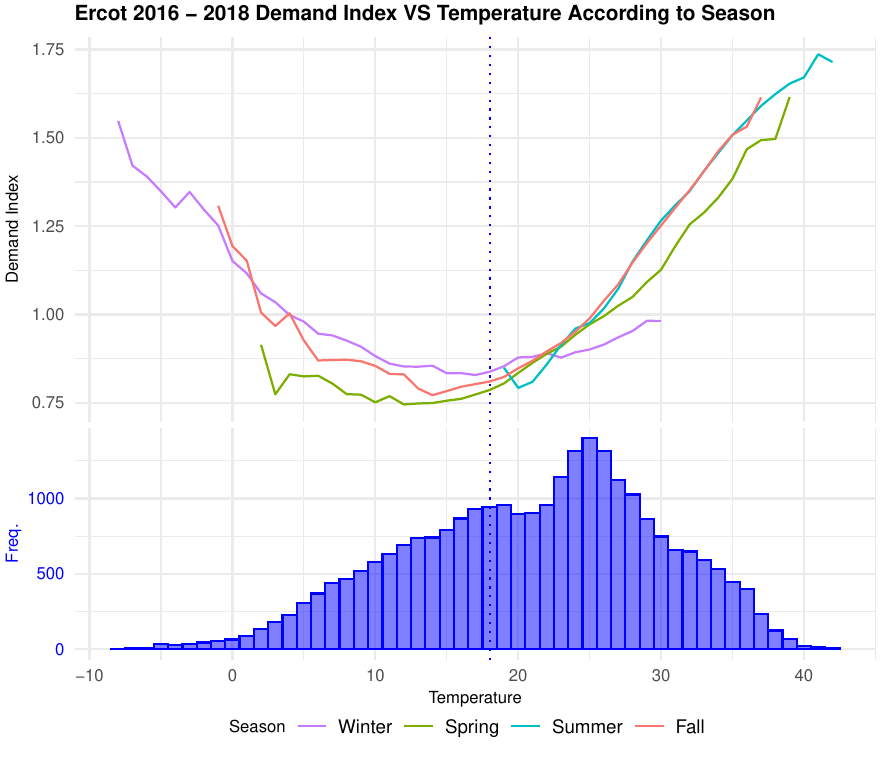}
\label{fig:er_season_temp}
\end{figure}

\renewcommand{\thefigure}{S2}
\begin{figure}[!h]
\centering
\caption{\footnotesize \textbf{ Demand in relation to temperature for each region in the Eastern interconnect, related to STAR Methods.} The panel of graphs show the average demand index value according to temperature and season per regional balancing authority region displayed above a frequency count of temperature. Region coverage is indicated relative to the eastern interconnect in the map. Panels display regions in order of coolest to warmest average temperature A through O.}
\includegraphics[trim=0cm 0cm 0cm 0.5cm, clip=true, width=0.9\textwidth]{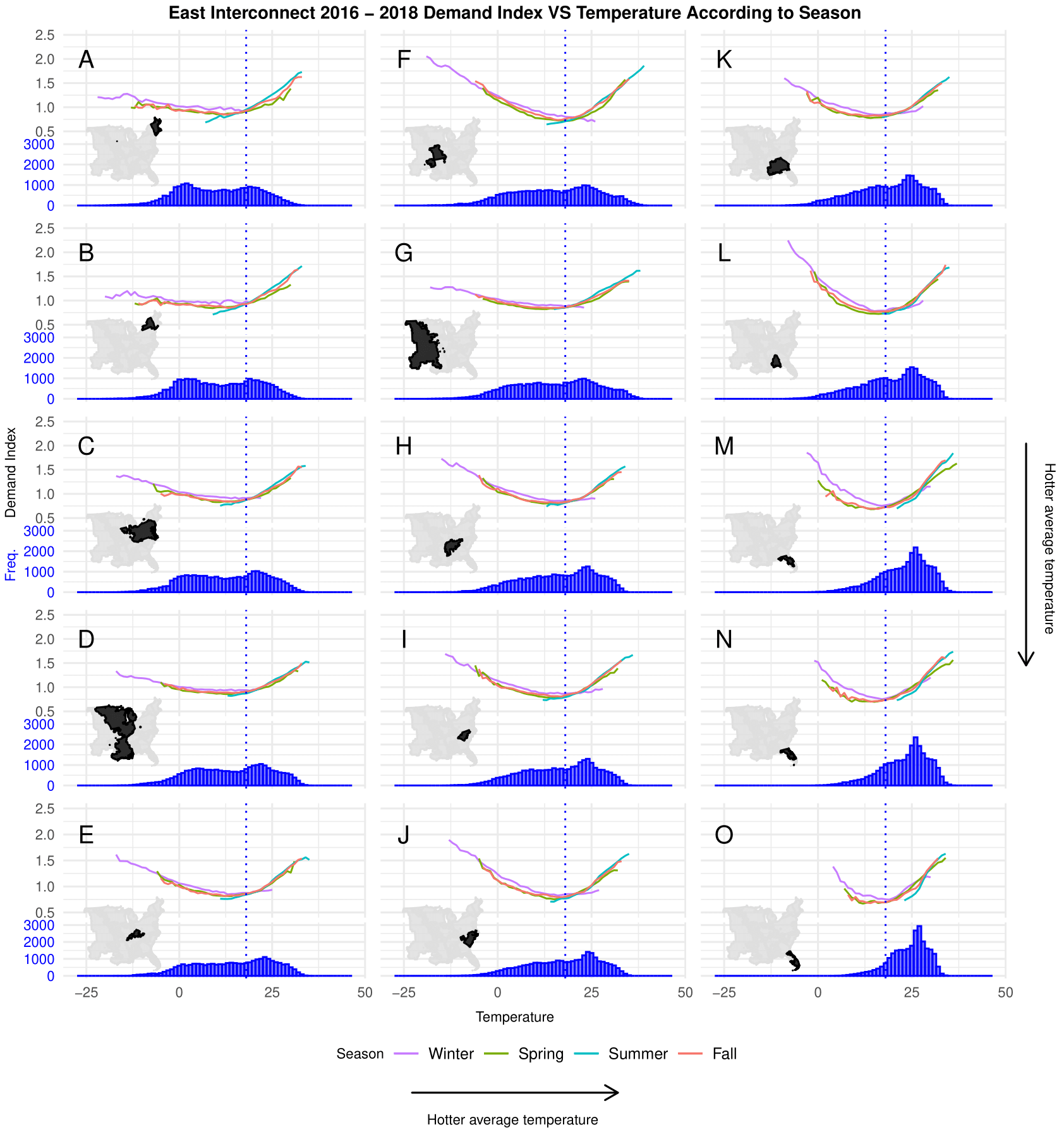}
\label{fig:e_season_temp}
\end{figure}

\renewcommand{\thefigure}{S3}
\begin{figure}[!h]
\centering
\caption{\footnotesize \textbf{Demand in relation to temperature for each region in the Western interconnect, related to STAR Methods.} The panel of graphs show the average demand index value according to temperature and season per regional balancing authority region displayed above a frequency count of temperature. Region coverage is indicated relative to the western interconnect in the map. Panels display regions in order of coolest to warmest average temperature A through O.}
\includegraphics[trim=0cm 0cm 0cm 0.5cm, clip=true, width=0.9\textwidth]{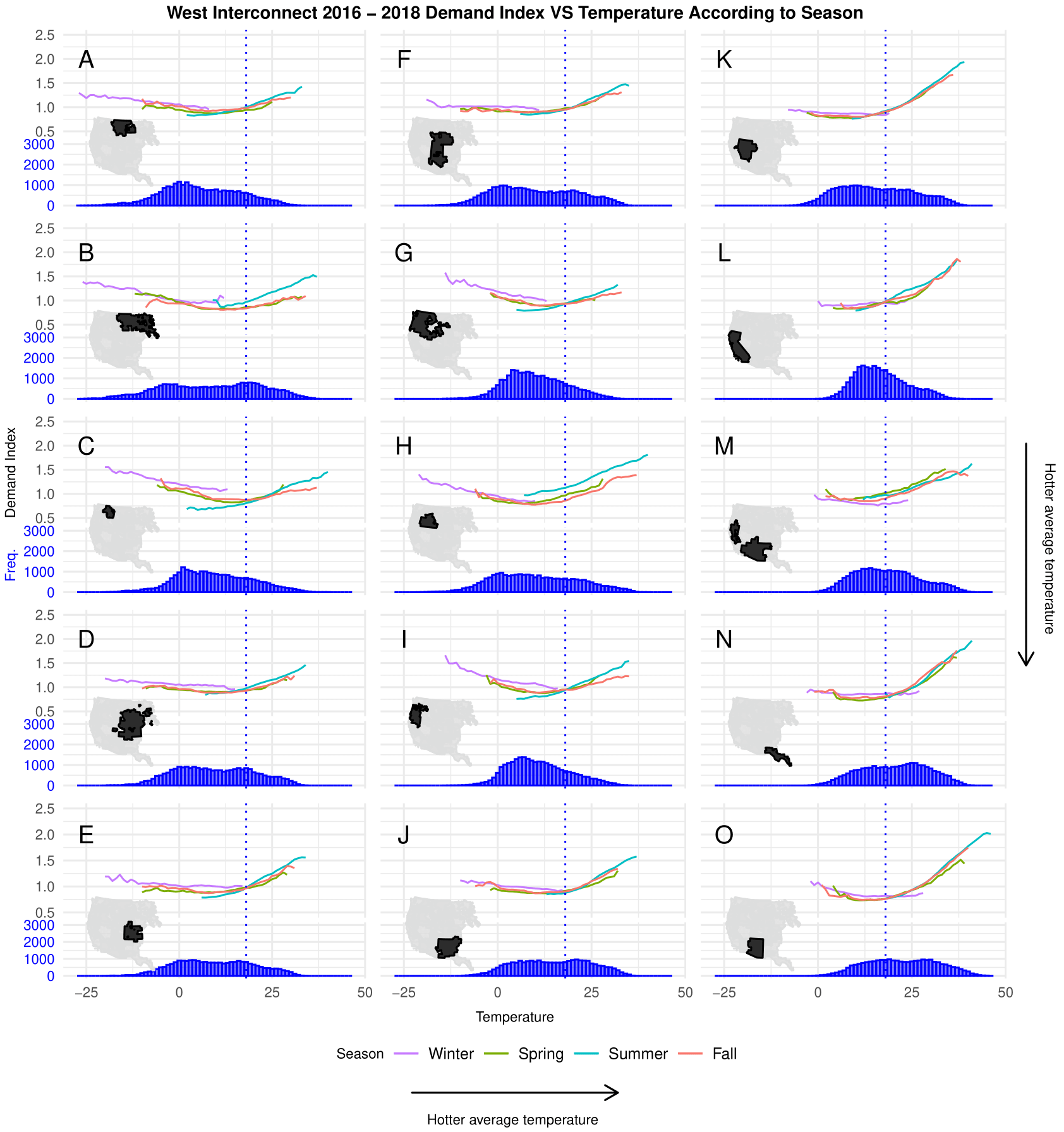}
\label{fig:w_season_temp}
\end{figure}

\renewcommand{\thefigure}{S4}
\begin{figure}[!h]
\centering
\caption{\footnotesize \textbf{ Demand in relation to heating and cooling degree hours for each region in the Eastern interconnect, related to STAR Methods and Figure 3.} The panel of graphs show the average demand index value according to value of cooling or heating degree hour is displayed per regional balancing authority region displayed above a frequency count of CDH or HDH (overlaid). Region coverage is indicated relative to the eastern interconnect in the map. Panels display regions in order of coolest to warmest average temperature A through O.}
\includegraphics[trim=0cm 0cm 0cm 0.5cm, clip=true, width=0.9\textwidth]{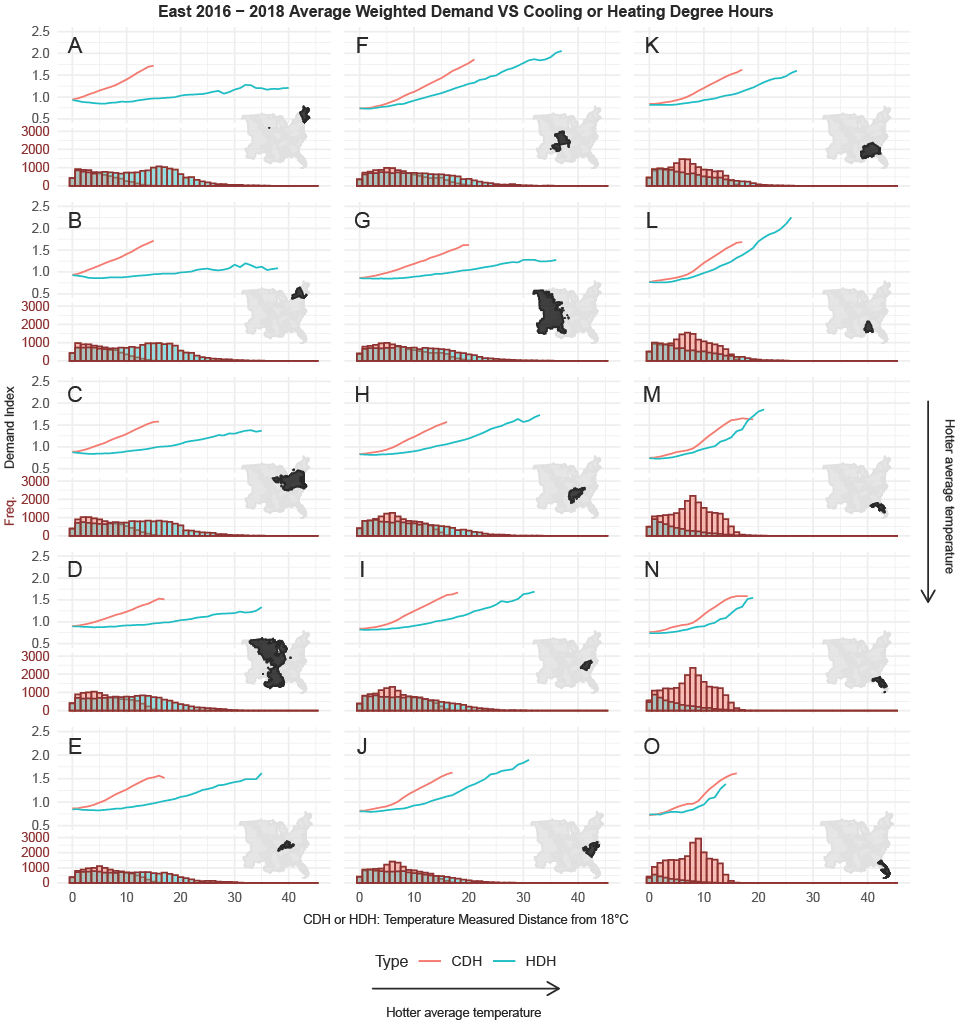}
\label{fig:e_cdhhdh_temp}
\end{figure}

\renewcommand{\thefigure}{S5}
\begin{figure}[!h]
\centering
\caption{\footnotesize \textbf{Demand in relation to heating and cooling degree hours for each region in the Western interconnect, related to STAR Methods and Figure 3.} The panel of graphs show the average demand index value according to value of cooling or heating degree hour per regional balancing authority region displayed above a frequency count of CDH and HDH (overlaid). Region coverage is indicated relative to the western interconnect in the map. Panels display regions in order of coolest to warmest average temperature A through O.}
\includegraphics[trim=0cm 0cm 0cm 0.5cm, clip=true, width=0.9\textwidth]{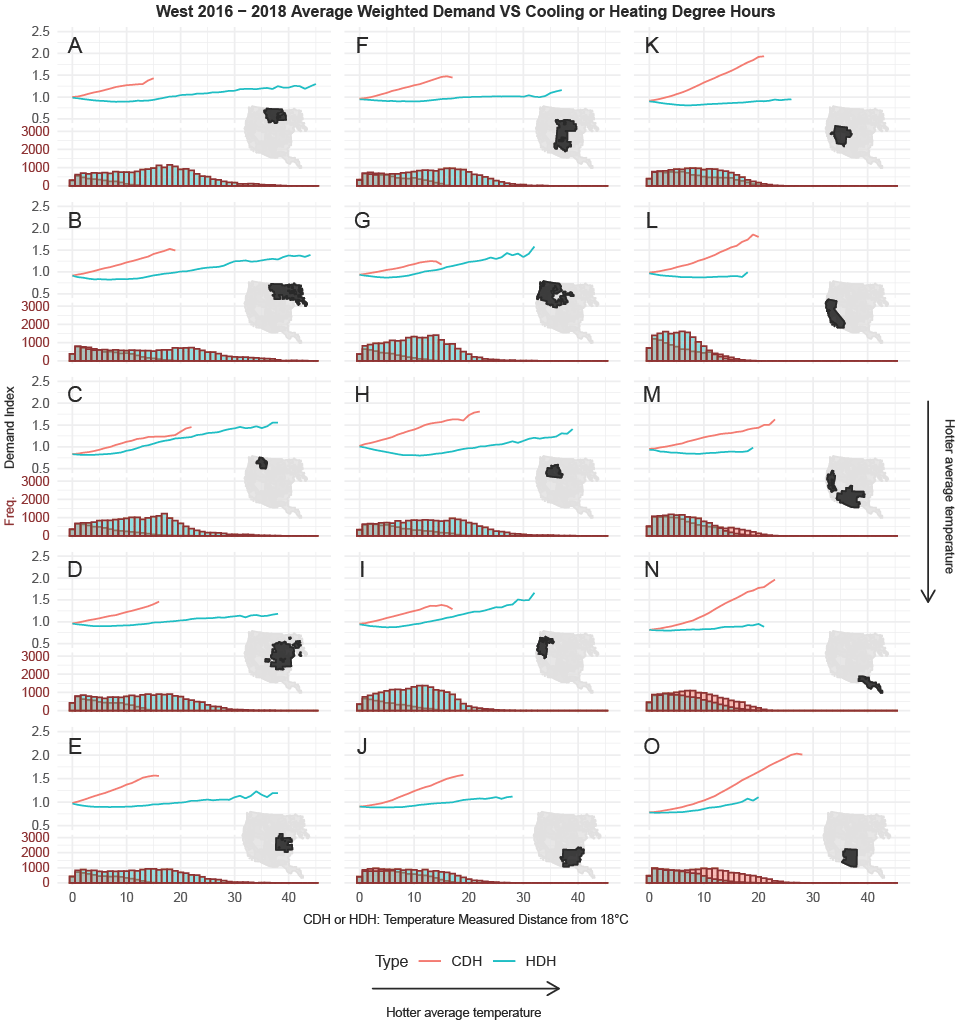}
\label{fig:w_cdhhdh_temp}
\end{figure}

\renewcommand{\thefigure}{S6}
\begin{figure}[h]
\centering
\caption{\scriptsize \textbf{Summary of regression models, related to STAR Methods.} The graphs summarize two regression specifications for each region, with and without an interaction between climate and weather outcomes. The top panel summarizes the Eastern interconnect and the bottom panel summarizes regression models for the Western interconnect. The first pair of points indicate the regression coefficients on CDH and HDH in the baseline model, with barely visible error bars to indicate uncertainty--the variables are highly statistically significant, even with Newey-West corrections to account for heteroskedasticity and autocorrelation. The next points with larger whiskers show the range of coefficients implied the model with weather-climate interactions, such that the weather response depends on mean CDH or mean HDH within the region. All models include controls for hour of year, day of week, and hour of day, which typically explain about two-thirds of demand variance in a region. The \% RMSE bars at the bottom report the share of \emph{remaining} variability explained by the weather and climate variables relative to a model that includes only the controls. The models with heterogeneous response have only slightly better predictions, and have no meaningful influence on other results. The size of the region (mean GW per hour) is indicated at the top of each panel, together with letter that corresponds to the region indicated in other figures above.}
\begin{tabular}{c}
\textbf{ Eastern Interconnect} \\
\includegraphics[trim=0cm 0cm 0cm 0.2cm, clip=true, width=0.75\textwidth]{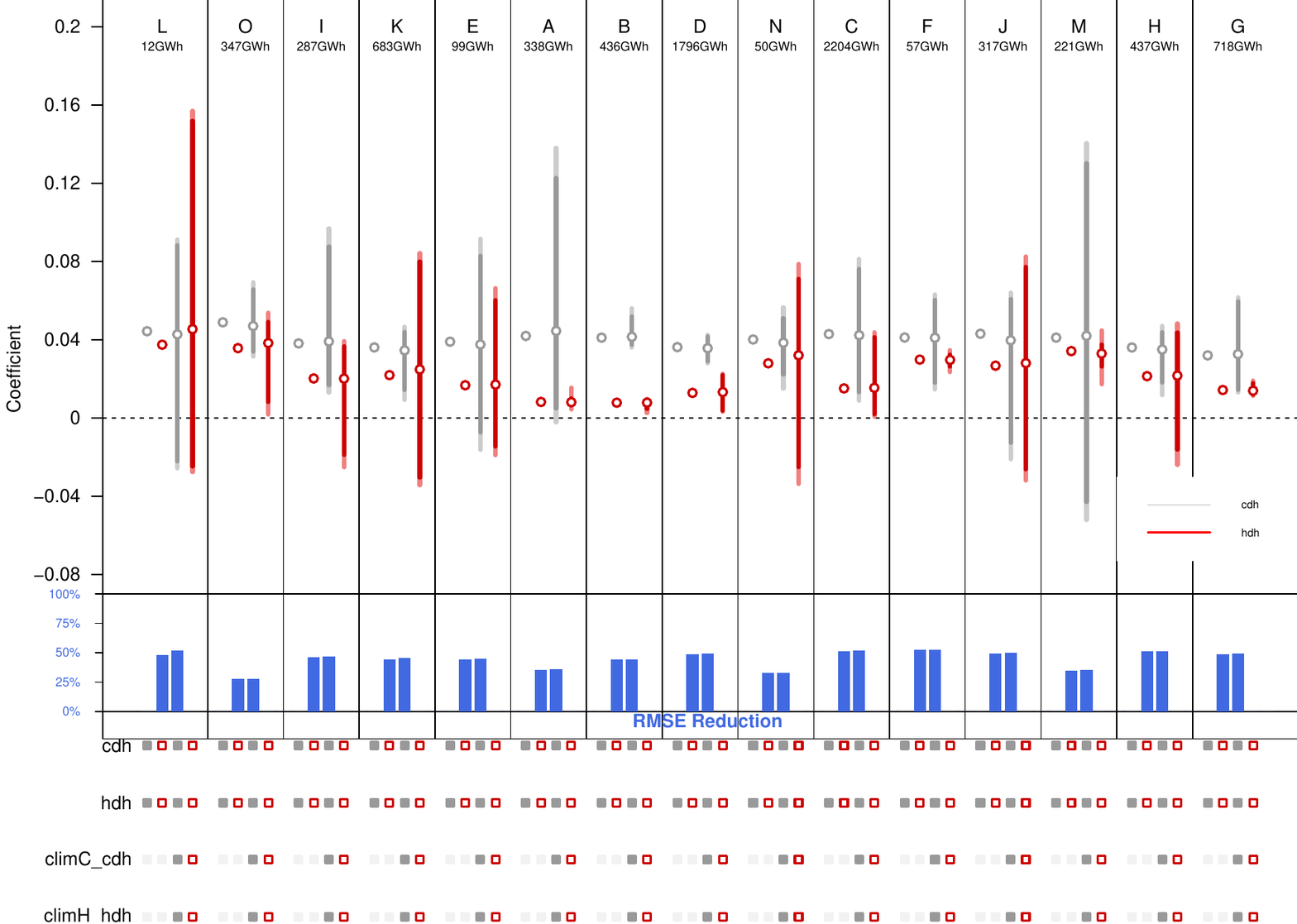} \\
\textbf{ Western Interconnect} \\
\includegraphics[trim=0cm 0cm 0cm 0.2cm, clip=true, width=0.75\textwidth]{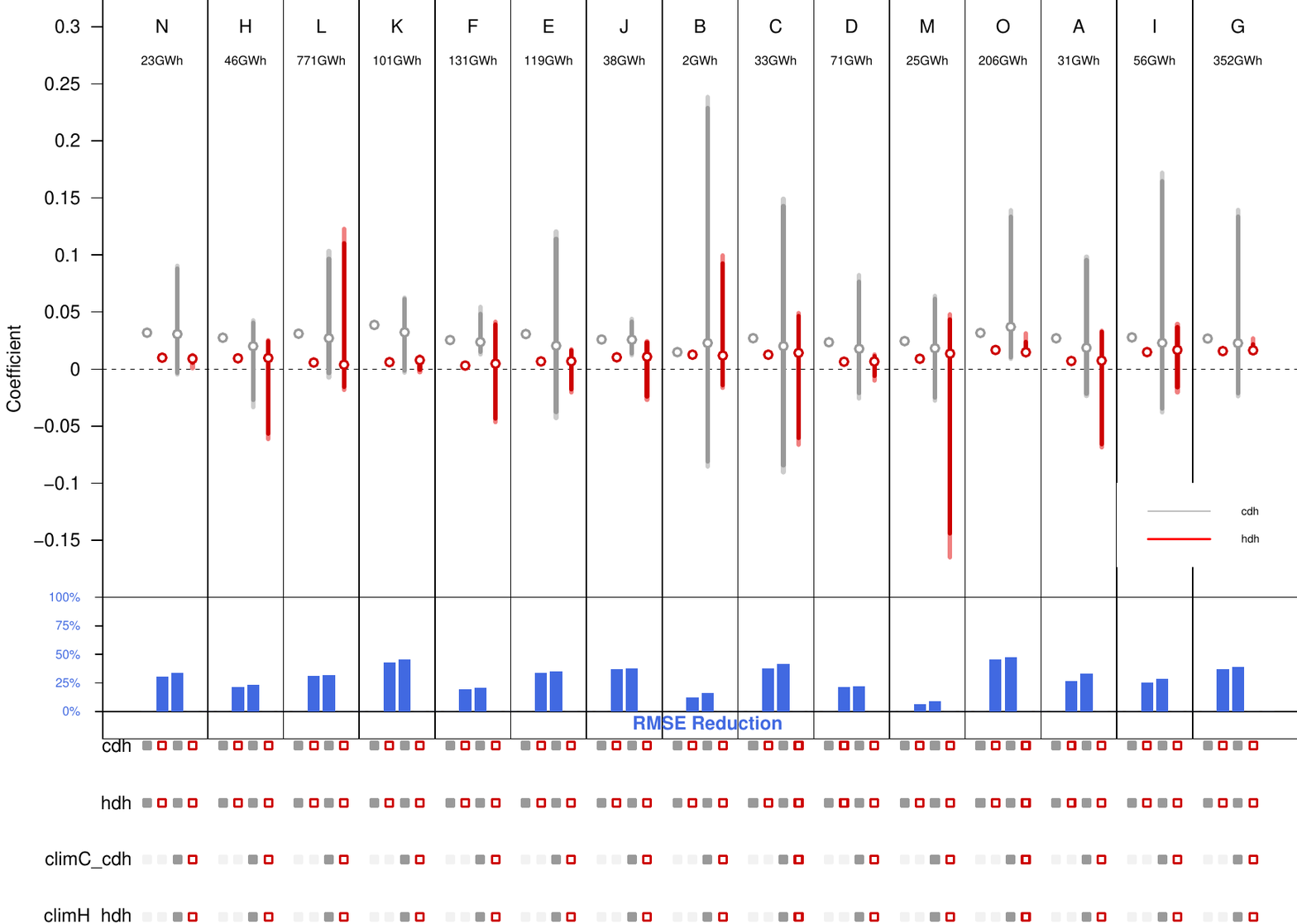}
\end{tabular}
\label{fig:reg_all_regions}
\end{figure}

\renewcommand{\thefigure}{S7}
\begin{figure}[!h]
\centering
\caption{\footnotesize \textbf{Average SD reduction for the contiguous United States ($\alpha=1$). Related to Figure 2 in Results.} The map shows the average daily reduction in standard deviation (SD) of demand if 100\% ($\alpha=1$) of potentially flexible load is actually flexible for each established balancing authority (BA) region. We use the average SD reduction of the two overlapped BAs as SD reduction of the overlapped areas. SD reduction ranges from 0.68 to 1.00 across the continental US.}
\includegraphics[trim=0cm 4cm 0cm 4cm, clip=true, width=0.9\textwidth]{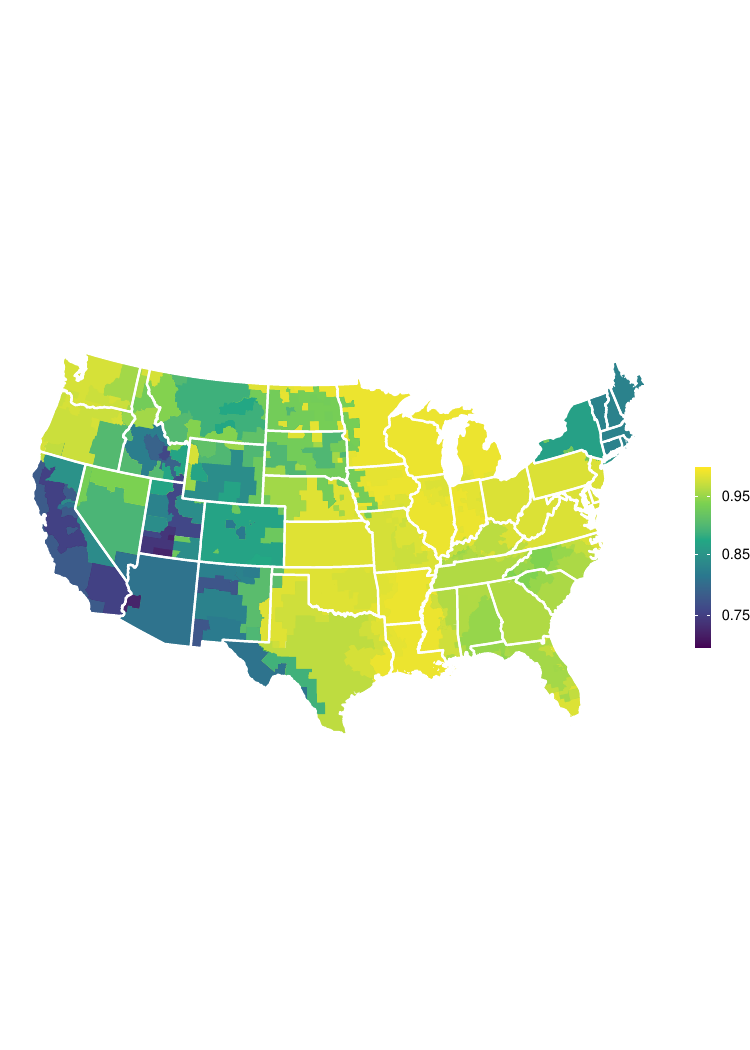}
\label{fig:sdred_map}
\end{figure}

\renewcommand{\thefigure}{S8}
\begin{figure}[t]
\centering
\caption{\footnotesize \textbf{Proportional reduction in overall load variability in different seasons. Related to Table 1 in Results.} Each colored line represents the average regional SD reduction in a season. The black line represents the average seasonal SD reduction value across all seasons 2016 - 2018.}. 
\includegraphics[trim=0cm 0cm 0cm 0.7cm, clip=true, width=0.9\textwidth]{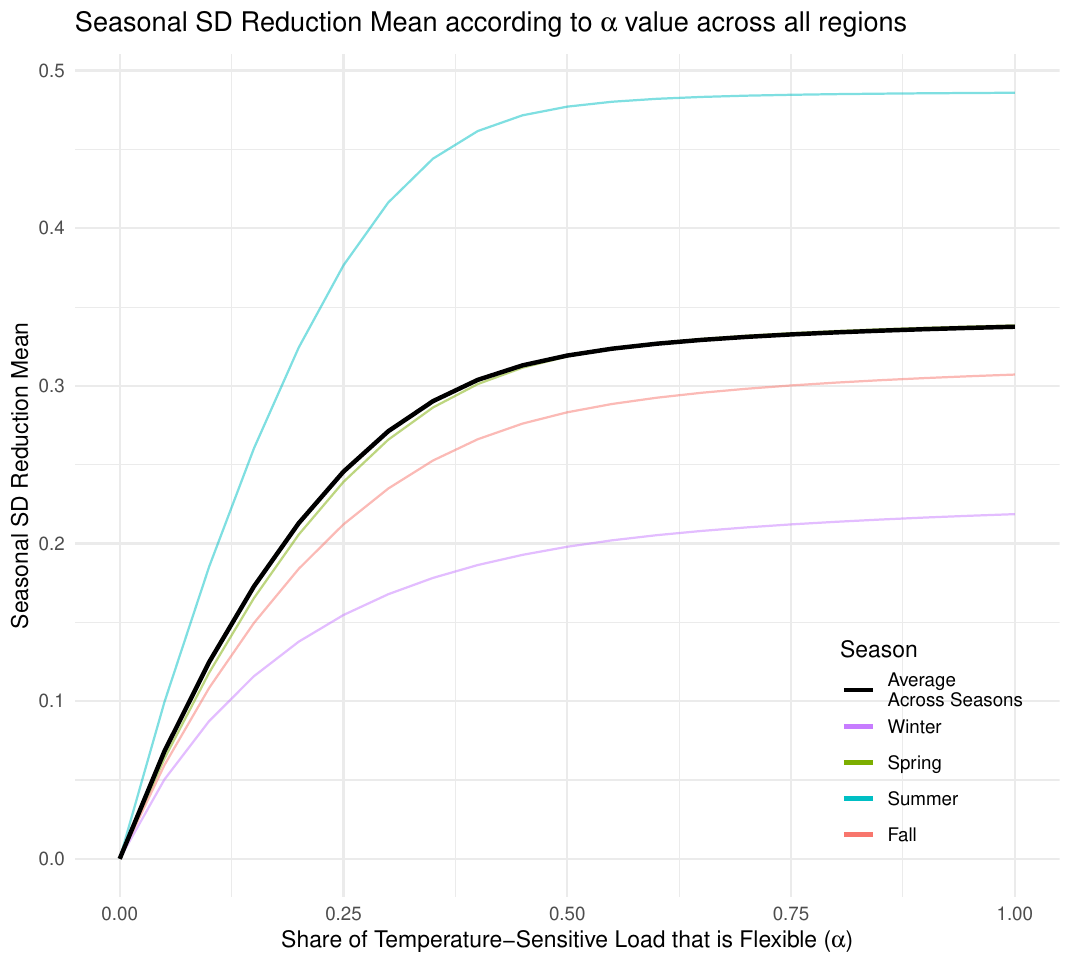}
\label{fig:seasonal_sd}
\end{figure}

\renewcommand{\thefigure}{S9}
\begin{figure}[t]
\centering
\caption{\footnotesize \textbf{The influence of demand flexibility and transmission on daily and overall base and peak load with and without climate interactions in the model. Related to STAR Methods.} The aqua colored bars show average values of daily peak and base load divided by the same-day mean (lighter shade) or overall (3-year) mean (darker shade). The purple bars indicate the same average values with climate interactions, normalized by the actual historical load (ie., mean CDH and HDH are interacted with hourly realizations of CDH and HDH). Whiskers mark the 1\textsuperscript{st} and 99\textsuperscript{th} percentiles of daily peak and base demand. Demand flexibility increases from left to right, where $\alpha=0$ is raw demand (left column), $\alpha=0.5$ is demand optimally flattened using half the temperature-sensitive load, and $\alpha=1$ is demand optimally flattened using all of the temperature-sensitive load. Transmission increases from top to bottom, where the first row assumes no connectivity between regions, the second row assumes perfect transmission within interconnects (East, West, ERCOT), and the last row assumes perfect transmission across the contiguous United States.}
\includegraphics[trim=0cm 0cm 0cm 0.7cm, clip=true, width=0.9\textwidth]{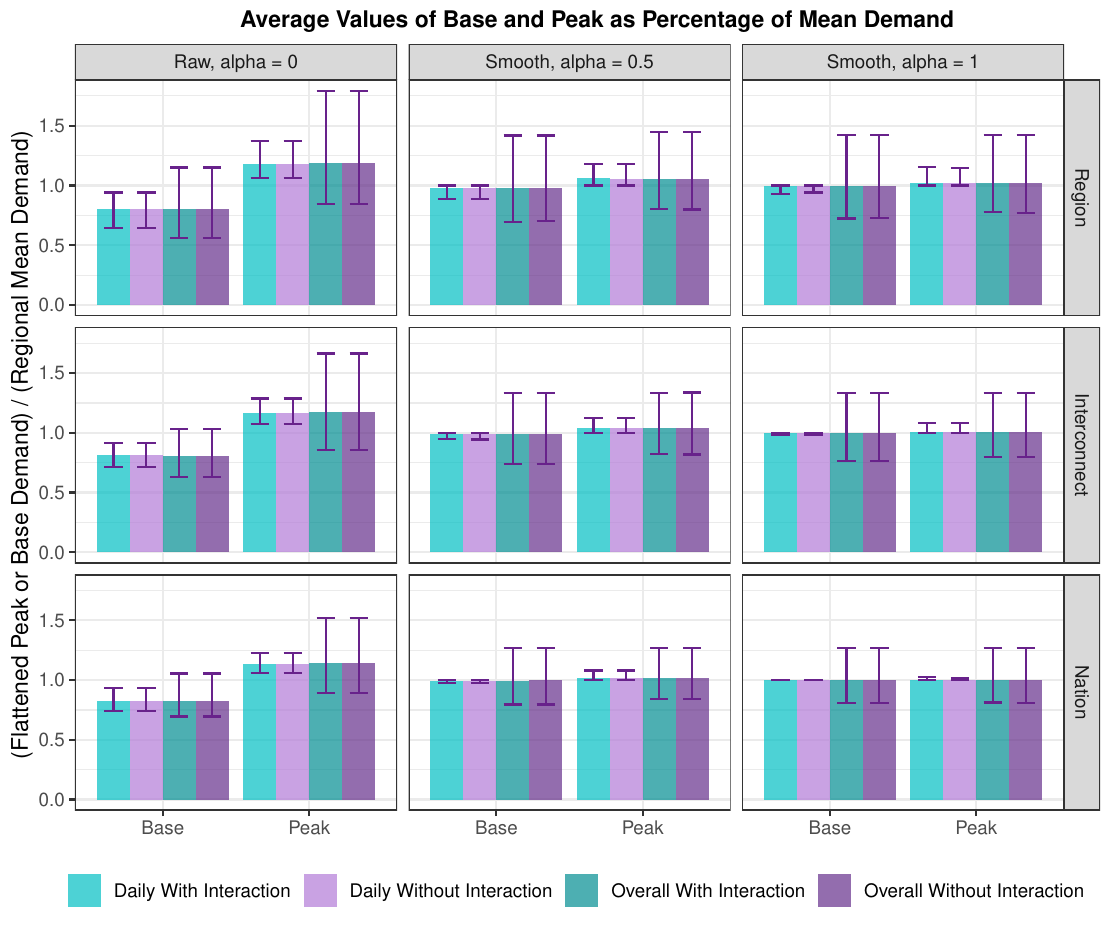}
\label{fig:trans_flex_ALL_interact}
\end{figure}

\renewcommand{\thetable}{S1}
\begin{table}[h!]
\centering
\caption{\textbf{Model comparisons for one region (L) in the Eastern interconnect, related to Regression Model in STAR Methods.}}
\begin{tabular}{l c c c} 
 \hline\hline
Independent  & \multicolumn{3}{c}{Models} \\
Variable &  Baseline &  Weather &  Weather+Climate\\ 
 & (a) & (b) & (c) \\ [0.5ex]  
 \hline
Heating Degree Hours  &  & 0.037  & 0.201  \\
         & & (0.0002) & (0.004)  \\
Mean HDH    & &  & -0.048  \\ 
          & &  & (0.001)  \\ 
Cooling Degree Hours  &  & 0.044 & -0.211   \\ 
         &  & (0.0003)  & (0.007)  \\ 
Mean CDH & &   & 0.059  \\
        & &  & (0.002)  \\
\hline
 s(Hour of Year) $\times$ s(Hour of Day) $\times$ I(Day of Week). & Yes & Yes & Yes \\
  Adjusted $R^2$ & 0.675 & 0.912 & \textbf{0.924}  \\
3 fold CV $R^2$ & 0.624 & 0.900 & \textbf{0.914} \\
 \hline\hline
\end{tabular}
\label{tab:reg_one}
\end{table}

\renewcommand{\thetable}{S2}
\begin{table}[h]
    \centering 
    \caption{\textbf{Changes in overall peak and base load, and overall variability, when demand is flattened within each day for different levels of $\alpha$ and regional aggregation. Related to Figure 4.}}
     \label{tab:smoothdem_stat_overall}
   \begin{tabular}{l c c c}
     \hline\hline
     Level of & Peak Load & Base Load & SD\\
     Connectivity & Reduction & Increase & Reduction\\ [0.5ex] 
     \hline 
     {$\alpha$ value} & 0 $\vert$ 0.25 $\vert$ 0.5 $\vert$ 1 & 0 $\vert$ 0.25 $\vert$ 0.5 $\vert$ 1 & 0 $\vert$ 0.25 $\vert$ 0.5 $\vert$ 1 \\
     \hline
     \multicolumn{4}{c}{-All Seasons-}\\
     \hline
     Regional BA & 0.0 $\vert$ 10.6 $\vert$ 16.6 $\vert$ 17.3 & 0.0 $\vert$ 16.7 $\vert$ 22.0 $\vert$ 27.1 & 0.0 $\vert$ 18.7 $\vert$ 24.7 $\vert$ 26.3\\ 
     Interconnect & 5.4 $\vert$ 15.7 $\vert$ 22.9 $\vert$ 23.1 & 5.9 $\vert$ 14.8 $\vert$ 18.4 $\vert$ 21.6 & 8.5 $\vert$ 28.8 $\vert$ 34.2 $\vert$ 35.0\\ 
     Nationwide & 10.0 $\vert$ 19.5 $\vert$ 25.3 $\vert$ 25.3 & 10.8 $\vert$ 23.0 $\vert$ 27.5 $\vert$ 28.6 & 14.5 $\vert$ 35.3 $\vert$ 39.6 $\vert$ 39.8\\ 
     \hline
     \multicolumn{4}{c}{-Winter—Dec/Jan/Feb-}\\
     \hline
     Regional BA & 0.0 $\vert$ 7.6 $\vert$ 10.8 $\vert$ 12.1 & 0.0 $\vert$ 10.7 $\vert$ 14.6 $\vert$ 17.8 & 0.0 $\vert$ 17.1 $\vert$ 23.1 $\vert$ 25.9\\ 
     Interconnect & 4.3 $\vert$ 11.2 $\vert$ 13.3 $\vert$ 14.5 & 2.6 $\vert$ 12.1 $\vert$ 16.4 $\vert$ 18.7 & 9.5 $\vert$ 26.4 $\vert$ 31.6 $\vert$ 33.7\\ 
     Nationwide & 8.8 $\vert$ 14.9 $\vert$ 14.9 $\vert$ 14.9 & 3.9 $\vert$ 13.5 $\vert$ 17.9 $\vert$ 20.0 & 21.7 $\vert$ 35.8 $\vert$ 37.4 $\vert$ 37.6\\
     \hline
     \multicolumn{4}{c}{-Summer—Jun/Jul/Aug-}\\
     \hline
     Regional BA & 0.0 $\vert$ 10.5 $\vert$ 18.6 $\vert$ 19.9 & 0.0 $\vert$ 18.9 $\vert$ 25.9 $\vert$ 31.1 & 0.0 $\vert$ 35.4 $\vert$ 47.8 $\vert$ 50.0\\ 
     Interconnect & 3.8 $\vert$ 14.3 $\vert$ 22.0 $\vert$ 22.2 & 3.6 $\vert$ 18.6 $\vert$ 25.2 $\vert$ 26.9 & 4.4 $\vert$ 44.9 $\vert$ 56.3 $\vert$ 56.9\\ 
     Nationwide & 8.2 $\vert$ 17.9 $\vert$ 23.9 $\vert$ 23.9 & 9.6 $\vert$ 23.2 $\vert$ 29.0 $\vert$ 31.5 & 11.9 $\vert$ 52.2 $\vert$ 61.5 $\vert$ 61.6\\ 
     \hline\hline
     \multicolumn{4}{p{6in}}{{\footnotesize \emph{Notes}: The table shows the average percent reduction in overall peak load, average percent increase in overall base load, average reduction in overall standard deviation of load when all ($\alpha=1$), half ($\alpha=0.5$), a quarter ($\alpha =0.25$) or none of temperature-sensitive load in each day is shiftable to another hour in the same day. This differs from the table in the main paper that considers \emph{daily} peak and base loads instead of \emph{overall} (3-year) peak and base loads. Baseline mean values are overall averages. These calculations are performed for the individual regions, when regions are pooled within each interconnect, and when all regions across the continental United States are pooled (Nationwide). The calculations are also broken out for Winter and Summer months.}}
    \end{tabular}
 \end{table}

\renewcommand{\thetable}{S3}
\begin{sidewaystable}[h]
    \centering 
    \caption{\textbf{Summary of 1\%, mean, 99\% of daily peak and base loads relative to the historic daily or overall mean, when demand is flattened within each day for different levels of $\alpha$ and regional aggregation,  under both present and +2$^\circ$C temperatures. Related to Results and Figure 4 in the main text.}}
     \label{tab:bargraphtable}
   \begin{tabular}{=l +c +c +c +c +c +c +c +c}
\hline\hline
Trans- & $\alpha$ & Temp. &  &  & & & & \\ 
mission & Value & Scenario & Base 1\% & Base Mean & Base 99\% & Peak 1\% & Peak Mean & Peak 99\% \\
\hline
\multicolumn{3}{c}{Baseline Mean Value} & D $\vert$ O & D $\vert$ O & D $\vert$ O & D $\vert$ O & D $\vert$ O & D $\vert$ O \\
\hline
\multirow{6}{*}{\begin{turn}{90} Region \end{turn}} & 0 & Present & 64.3 $\vert$ 56.4 & 80.7 $\vert$ 80.4 & 94.1 $\vert$ 115.2 & 106.5 $\vert$ 84.3 & 118.2 $\vert$ 118.6 & 137.2 $\vert$ 179.4 \\ 
\rowstyle{\color{BrickRed}} & 0 & +2$^{\circ}$C & 66.7 $\vert$ 55.2 & 81.2 $\vert$ 81.1 & 92.9 $\vert$ 116.2 & 104.3 $\vert$ 84.1 & 121.9 $\vert$ 122.5 & 148.7 $\vert$ 191.0 \\ 
& 0.5 & Present & 88.7 $\vert$ 70.4 & 97.8 $\vert$ 98.0 & 100.0 $\vert$ 141.8 & 100.0 $\vert$ 80.0 & 105.9 $\vert$ 105.5 & 118.0 $\vert$ 144.9 \\ 
\rowstyle{\color{BrickRed}} & 0.5 & +2$^{\circ}$C & 87.5 $\vert$ 70.0 & 100.0 $\vert$ 100.4 & 110.5 $\vert$ 149.3 & 95.0 $\vert$ 80.1 & 107.9 $\vert$ 107.7 & 121.3 $\vert$ 151.2 \\ 
& 1 & Present & 94.1 $\vert$ 72.9 & 99.5 $\vert$ 99.6 & 100.0 $\vert$ 142.0 & 100.0 $\vert$ 77.0 & 102.2 $\vert$ 102.0 & 114.7 $\vert$ 142.1 \\
\rowstyle{\color{BrickRed}} & 1 & +2$^{\circ}$C & 93.0 $\vert$ 72.7 & 101.7 $\vert$ 102.0 & 110.5 $\vert$ 149.4 & 94.4 $\vert$ 76.9 & 104.3 $\vert$ 104.2 & 114.9 $\vert$ 149.4 \\ 
\hline
\multirow{6}{*}{\begin{turn}{90} Interconnect \end{turn}} & 0 & Present & 71.3 $\vert$ 63.0 & 81.1 $\vert$ 80.8 & 91.8 $\vert$ 103.4 & 107.3 $\vert$ 85.7 & 116.6 $\vert$ 117.0 & 128.8 $\vert$ 166.4 \\ 
\rowstyle{\color{BrickRed}} & 0 & +2$^{\circ}$C & 72.9 $\vert$ 61.2 & 81.8 $\vert$ 81.7 & 89.8 $\vert$ 108.7 & 104.7 $\vert$ 87.4 & 120.6 $\vert$ 121.3 & 137.4 $\vert$ 177.7 \\  
& 0.5 & Present & 94.4 $\vert$ 74.0 & 98.9 $\vert$ 99.0 & 100.0 $\vert$ 133.4 & 100.0 $\vert$ 81.9 & 104.2 $\vert$ 103.9 & 112.6 $\vert$ 133.8 \\
\rowstyle{\color{BrickRed}} & 0.5 & +2$^{\circ}$C & 94.0 $\vert$ 73.8 & 101.4 $\vert$ 101.7 & 108.3 $\vert$ 142.2 & 96.8 $\vert$ 82.4 & 106.5 $\vert$ 106.5 & 114.6 $\vert$ 142.2 \\  
& 1 & Present & 98.6 $\vert$ 76.6 & 99.9 $\vert$ 99.9 & 100.0 $\vert$ 133.4 & 100.0 $\vert$ 79.7 & 101.1 $\vert$ 101.0 & 108.4 $\vert$ 133.4 \\  
\rowstyle{\color{BrickRed}} & 1 & +2$^{\circ}$C & 95.8 $\vert$ 76.6 & 102.4 $\vert$ 102.6 & 108.3 $\vert$ 142.2 & 95.9 $\vert$ 79.5 & 103.5 $\vert$ 103.6 & 108.3 $\vert$ 142.2 \\  
\hline
\multirow{6}{*}{\begin{turn}{90} Nation \end{turn}} & 0 & Present & 74.3 $\vert$ 69.6 & 82.7 $\vert$ 82.4 & 93.2 $\vert$ 105.6 & 106.1 $\vert$ 89.5 & 113.8 $\vert$ 114.1 & 122.8 $\vert$ 152.1 \\  
\rowstyle{\color{BrickRed}} & 0 & +2$^{\circ}$C & 77.5 $\vert$ 68.9 & 83.5 $\vert$ 83.4 & 90.9 $\vert$ 106.0 & 103.8 $\vert$ 90.6 & 117.8 $\vert$ 118.4 & 132.3 $\vert$ 163.9 \\  
& 0.5 & Present & 97.7 $\vert$ 79.6 & 99.6 $\vert$ 99.7 & 100.0 $\vert$ 127.1 & 100.0 $\vert$ 84.0 & 101.7 $\vert$ 101.5 & 108.0 $\vert$ 127.1 \\  
\rowstyle{\color{BrickRed}} & 0.5 & +2$^{\circ}$C & 97.1 $\vert$ 80.0 & 102.1 $\vert$ 102.4 & 107.6 $\vert$ 136.7 & 97.2 $\vert$ 84.5 & 104.1 $\vert$ 104.2 & 110.2 $\vert$ 136.7 \\  
& 1 & Present & 99.9 $\vert$ 80.6 & 100.0 $\vert$ 100.0 & 100.0 $\vert$ 127.1 & 100.0 $\vert$ 81.1 & 100.1 $\vert$ 100.0 & 102.0 $\vert$ 127.1 \\ 
\rowstyle{\color{BrickRed}} & 1 & +2$^{\circ}$C & 97.2 $\vert$ 81.3 & 102.5 $\vert$ 102.7 & 107.6 $\vert$ 136.7 & 97.2 $\vert$ 81.5 & 102.5 $\vert$ 102.7 & 107.6 $\vert$ 136.7 \\ 
\hline\hline
\multicolumn{9}{p{7.9in}}{{\footnotesize \emph{Notes}: The table shows the average, 1\textsuperscript{st} percentile, and 99\textsuperscript{th} percentile peak and base values in relation to the historical daily (D) or overall (O) mean demand when all ($\alpha=1$), half ($\alpha=0.5$), or none of temperature-sensitive load in each day is shiftable ($\alpha$ = 0), and for all levels aggregation (within each region, within interconnect, or nationwide).}}
    \end{tabular}
 \end{sidewaystable}

\renewcommand{\thetable}{S4}
\begin{sidewaystable}[h]
    \centering 
    \caption{\textbf{Changes in daily peak and base load, and daily variability, when demand is flattened within each day for different levels of $\alpha$ and regional aggregation, using models with climate-weather interactions. Related to STAR Methods.}}
    \label{tab:smoothdem_stat_daily_interact}
   \begin{tabular}{l c c c c}
     \hline\hline
     Level of & Peak Load & Base Load & SD & Share of\\
     Connectivity & Reduction & Increase & Reduction & Flattenable Days\\ [0.5ex] 
     \hline 
     {$\alpha$ value} & 0 $\vert$ 0.25 $\vert$ 0.5 $\vert$ 1 & 0 $\vert$ 0.25 $\vert$ 0.5 $\vert$ 1 & 0 $\vert$ 0.25 $\vert$ 0.5 $\vert$ 1 & 0 $\vert$ 0.25 $\vert$ 0.5 $\vert$ 1 \\
     \hline
     \multicolumn{5}{c}{-All Seasons-}\\
     \hline
     Regional BA & 0.0 $\vert$ 5.4 $\vert$ 10.0 $\vert$ 13.0 & 0.0 $\vert$ 16.4 $\vert$ 22.1 $\vert$ 24.3 & 0.0 $\vert$ 47.7 $\vert$ 75.8 $\vert$ 90.9 & 0.0 $\vert$ 0.7 $\vert$ 16.8 $\vert$ 59.2\\ 
     Interconnect & 0.9 $\vert$ 6.5 $\vert$ 11.3 $\vert$ 13.9 & 0.6 $\vert$ 17.8 $\vert$ 23.1 $\vert$ 24.3 & 4.6 $\vert$ 55.9 $\vert$ 85.5 $\vert$ 97.2 & 0.0 $\vert$ 0.4 $\vert$ 23.1 $\vert$ 69.6\\ 
     Nationwide & 2.5 $\vert$ 8.0 $\vert$ 12.6 $\vert$ 14.1 & 2.0 $\vert$ 18.8 $\vert$ 23.2 $\vert$ 23.7 & 12.7 $\vert$ 66.3 $\vert$ 93.8 $\vert$ 99.8 & 0.0 $\vert$ 1.0 $\vert$ 40.2 $\vert$ 95.0\\ 
     \hline
     \multicolumn{5}{c}{-Winter—Dec/Jan/Feb-}\\
     \hline
     Regional BA & 0.0 $\vert$ 4.3 $\vert$ 7.6 $\vert$ 10.3 & 0.0 $\vert$ 12.4 $\vert$ 15.8 $\vert$ 17.5 & 0.0 $\vert$ 51.3 $\vert$ 75.1 $\vert$ 90.3 & 0.0 $\vert$ 2.5 $\vert$ 22.0 $\vert$ 55.1\\ 
     Interconnect & 1.1 $\vert$ 4.9 $\vert$ 8.0 $\vert$ 11.0 & 0.8 $\vert$ 13.5 $\vert$ 16.8 $\vert$ 17.8 & 7.7 $\vert$ 58.1 $\vert$ 82.1 $\vert$ 96.4 & 0.0 $\vert$ 1.5 $\vert$ 18.6 $\vert$ 58.4\\ 
     Nationwide & 3.5 $\vert$ 7.5 $\vert$ 10.3 $\vert$ 11.4 & 2.6 $\vert$ 15.1 $\vert$ 16.9 $\vert$ 17.2 & 21.1 $\vert$ 76.8 $\vert$ 95.0 $\vert$ 99.8 & 0.0 $\vert$ 4.1 $\vert$ 49.8 $\vert$ 95.5\\ 
     
     \hline
     \multicolumn{5}{c}{-Summer—Jun/Jul/Aug-}\\
     \hline
     Regional BA & 0.0 $\vert$ 8.1 $\vert$ 15.3 $\vert$ 18.0 & 0.0 $\vert$ 22.9 $\vert$ 30.5 $\vert$ 32.1 & 0.0 $\vert$ 53.8 $\vert$ 87.4 $\vert$ 97.6 & 0.0 $\vert$ 0.0 $\vert$ 28.0 $\vert$ 85.8\\ 
     Interconnect & 0.5 $\vert$ 9.5 $\vert$ 17.4 $\vert$ 18.6 & 0.5 $\vert$ 24.2 $\vert$ 30.8 $\vert$ 31.2 & 1.5 $\vert$ 61.3 $\vert$ 96.3 $\vert$ 100.0 & 0.0 $\vert$ 0.0 $\vert$ 49.8 $\vert$ 98.6\\ 
     Nationwide & 1.5 $\vert$ 10.3 $\vert$ 17.6 $\vert$ 18.0 & 1.8 $\vert$ 24.9 $\vert$ 30.8 $\vert$ 30.9 & 5.9 $\vert$ 65.9 $\vert$ 98.9 $\vert$ 100.0 & 0.0 $\vert$ 0.0 $\vert$ 75.3 $\vert$ 100.0\\ 
     
     \hline\hline
     \multicolumn{5}{p{7.5in}}{{\footnotesize \emph{Notes}: The table shows the average percent reduction in daily peak load, average percent increase in daily base load, average reduction in daily standard deviation of load when all ($\alpha=1$), half ($\alpha=0.5$), or a quarter ($\alpha =0.25$) of temperature-sensitive load in each day is shiftable to another hour in the same day using the model with climate interactions in the weather response to HDH and CDH. This differs from Table 1 in the main paper, which does not include these interactions. Baseline load values used were daily averages. These calculations are performed for the individual regions, when regions are pooled within each interconnect, and when all regions across the continental United States are pooled (Nationwide). The calculations are also broken out for Winter and Summer months.}}
    \end{tabular}
 \end{sidewaystable}

\renewcommand{\thetable}{S5}
\begin{table}[h]
    \centering 
    \caption{\textbf{Changes in overall peak and base load, and overall variability, when demand is flattened within each day for different levels of $\alpha$ and regional aggregation, and climate-weather interactions are used in the regression model that links demand to weather. Related to STAR Methods.}}
     \label{tab:smoothdem_stat_overall_interact}
   \begin{tabular}{l c c c}
     \hline\hline
     Level of & Peak Load & Base Load & SD\\
     Connectivity & Reduction & Increase & Reduction\\ [0.5ex] 
     \hline 
     {$\alpha$ value} & 0 $\vert$ 0.25 $\vert$ 0.5 $\vert$ 1 & 0 $\vert$ 0.25 $\vert$ 0.5 $\vert$ 1 & 0 $\vert$ 0.25 $\vert$ 0.5 $\vert$ 1 \\
     \hline
     \multicolumn{4}{c}{-All Seasons-}\\
     \hline
     Regional BA & 0.0 $\vert$ 10.6 $\vert$ 16.5 $\vert$ 17.3 & 0.0 $\vert$ 16.2 $\vert$ 21.3 $\vert$ 26.3 & 0.0 $\vert$ 18.6 $\vert$ 24.7 $\vert$ 26.2\\ 
     Interconnect & 5.4 $\vert$ 15.8 $\vert$ 23.1 $\vert$ 23.1 & 5.9 $\vert$ 14.8 $\vert$ 18.4 $\vert$ 21.8 & 8.5 $\vert$ 29.0 $\vert$ 34.3 $\vert$ 35.0\\ 
     Nationwide & 10.0 $\vert$ 19.5 $\vert$ 25.3 $\vert$ 25.3 & 10.8 $\vert$ 22.8 $\vert$ 27.4 $\vert$ 28.6 & 14.5 $\vert$ 35.2 $\vert$ 39.6 $\vert$ 39.8\\ 
     \hline
     \multicolumn{4}{c}{-Winter—Dec/Jan/Feb-}\\
     \hline
     Regional BA & 0.0 $\vert$ 7.7 $\vert$ 10.8 $\vert$ 12.1 & 0.0 $\vert$ 9.8 $\vert$ 13.7 $\vert$ 17.0 & 0.0 $\vert$ 16.7 $\vert$ 22.7 $\vert$ 25.8\\ 
     Interconnect & 4.3 $\vert$ 11.3 $\vert$ 13.4 $\vert$ 14.5 & 2.6 $\vert$ 12.1 $\vert$ 16.4 $\vert$ 18.8 & 9.5 $\vert$ 26.6 $\vert$ 31.8 $\vert$ 33.8\\ 
     Nationwide & 8.8 $\vert$ 14.9 $\vert$ 14.9 $\vert$ 14.9 & 3.9 $\vert$ 13.2 $\vert$ 17.6 $\vert$ 19.9 & 21.7 $\vert$ 35.6 $\vert$ 37.4 $\vert$ 37.6 \\ 
     \hline
     \multicolumn{4}{c}{-Summer—Jun/Jul/Aug-}\\
     \hline
     Regional BA & 0.0 $\vert$ 10.4 $\vert$ 18.6 $\vert$ 19.9 & 0.0 $\vert$ 18.1 $\vert$ 25.0 $\vert$ 30.3 & 0.0 $\vert$ 35.6 $\vert$ 48.3 $\vert$ 50.0\\ 
     Interconnect & 3.8 $\vert$ 14.3 $\vert$ 22.2 $\vert$ 22.2 & 3.6 $\vert$ 18.8 $\vert$ 25.4 $\vert$ 27.0 & 4.4 $\vert$ 45.3 $\vert$ 56.6 $\vert$ 56.9\\ 
     Nationwide & 8.2 $\vert$ 17.9 $\vert$ 23.9 $\vert$ 23.9 & 9.6 $\vert$ 23.2 $\vert$ 29.0 $\vert$ 31.5 & 11.9 $\vert$ 52.2 $\vert$ 61.5 $\vert$ 61.6\\ 
     \hline\hline
     \multicolumn{4}{p{6in}}{{\footnotesize \emph{Notes}: The table shows the average percent reduction in overall peak load, average percent increase in overall base load, average reduction in overall standard deviation of load when all ($\alpha=1$), half ($\alpha=0.5$), or a quarter ($\alpha =0.25$) of temperature-sensitive load in each day is shiftable to another hour in the same day using the model with climate interactions.  This differs from the table in the main paper that considers \emph{daily} peak and base loads instead of \emph{overall} (3-year) peak and base loads, \emph{and} it uses a regression model with climate-HDH/CDH interactions instead of just HDH and CDH. Baseline load values used were overall averages. These calculations are performed for the individual regions, when regions are pooled within each interconnect, and when all regions across the continental United States are pooled (Nationwide). The calculations are also broken out for Winter and Summer months.}}
    \end{tabular}
 \end{table}

\renewcommand{\thetable}{S6}
\begin{sidewaystable}[h]
    \centering 
    \caption{\textbf{Changes in daily peak and base load, and daily variability, when demand has been shifted +2$^\circ$C climate change and then flattened within each day for different levels of $\alpha$ and regional aggregation. Related to Figure 4.}}
     \label{tab:smoothdem_stat_daily_2C}
   \begin{tabular}{l c c c c}
    \hline\hline
     Level of & Peak Load & Base Load & SD & Share of\\
     Connectivity & Reduction & Increase & Reduction & Flattenable Days\\ [0.5ex] 
     \hline 
     {$\alpha$ value} & 0 $\vert$ 0.25 $\vert$ 0.5 $\vert$ 1 & 0 $\vert$ 0.25 $\vert$ 0.5 $\vert$ 1 & 0 $\vert$ 0.25 $\vert$ 0.5 $\vert$ 1 & 0 $\vert$ 0.25 $\vert$ 0.5 $\vert$ 1 \\
     \hline
     \multicolumn{5}{c}{-All Seasons-}\\
     \hline
     Regional BA & 0.0 $\vert$ 3.3 $\vert$ 8.5 $\vert$ 11.6 & 0.0 $\vert$ 19.3 $\vert$ 25.3 $\vert$ 27.4 & 0.0 $\vert$ 45.7 $\vert$ 75.8 $\vert$ 92.1 & 0.0 $\vert$ 0.5 $\vert$ 22.3 $\vert$ 62.1\\ 
     Interconnect & 0.9 $\vert$ 6.3 $\vert$ 11.4 $\vert$ 13.9 & 0.6 $\vert$ 17.6 $\vert$ 23.1 $\vert$ 24.3 & 4.6 $\vert$ 53.5 $\vert$ 84.8 $\vert$ 97.1 & 0.0 $\vert$ 0.3 $\vert$ 29.5 $\vert$ 71.4\\ 
     Nationwide & 2.5 $\vert$ 7.9 $\vert$ 12.8 $\vert$ 14.1 & 2.0 $\vert$ 18.8 $\vert$ 23.3 $\vert$ 23.7 & 12.7 $\vert$ 64.7 $\vert$ 94.0 $\vert$ 99.9 & 0.0 $\vert$ 0.5 $\vert$ 46.3 $\vert$ 97.3\\  
     \hline
     \multicolumn{5}{c}{-Winter—Dec/Jan/Feb-}\\
     \hline
     Regional BA & 0.0 $\vert$ 5.1 $\vert$ 8.5 $\vert$ 11.7 & 0.0 $\vert$ 10.7 $\vert$ 14.6 $\vert$ 16.4 & 0.0 $\vert$ 48.0 $\vert$ 74.2 $\vert$ 91.2 & 0.0 $\vert$ 1.8 $\vert$ 17.5 $\vert$ 56.8\\ 
     Interconnect & 1.1 $\vert$ 4.5 $\vert$ 7.6 $\vert$ 10.8 & 0.8 $\vert$ 12.8 $\vert$ 16.5 $\vert$ 17.8 & 7.7 $\vert$ 53.4 $\vert$ 79.7 $\vert$ 95.7 & 0.0 $\vert$ 1.4 $\vert$ 16.5 $\vert$ 56.1\\ 
     Nationwide & 3.5 $\vert$ 7.2 $\vert$ 10.2 $\vert$ 11.5 & 2.6 $\vert$ 14.7 $\vert$ 16.9 $\vert$ 17.2 & 21.1 $\vert$ 73.5 $\vert$ 94.4 $\vert$ 99.9 & 0.0 $\vert$ 1.9 $\vert$ 43.4 $\vert$ 96.6\\  
     
     \hline
     \multicolumn{5}{c}{-Summer—Jun/Jul/Aug-}\\
     \hline
     Regional BA & 0.0 $\vert$ 3.0 $\vert$ 10.9 $\vert$ 13.2 & 0.0 $\vert$ 31.4 $\vert$ 38.9 $\vert$ 40.4 & 0.0 $\vert$ 55.4 $\vert$ 88.8 $\vert$ 98.1 & 0.0 $\vert$ 0.1 $\vert$ 46.2 $\vert$ 87.8 \\ 
     Interconnect & 0.5 $\vert$ 9.9 $\vert$ 17.6 $\vert$ 18.6 & 0.5 $\vert$ 24.8 $\vert$ 30.9 $\vert$ 31.2 & 1.5 $\vert$ 63.7 $\vert$ 96.9 $\vert$ 100.0 & 0.0 $\vert$ 0.0 $\vert$ 62.4 $\vert$ 99.2\\ 
     Nationwide & 1.5 $\vert$ 10.9 $\vert$ 17.8 $\vert$ 18.0 & 1.8 $\vert$ 25.8 $\vert$ 30.9 $\vert$ 30.9 & 5.9 $\vert$ 69.5 $\vert$ 99.5 $\vert$ 100.0 & 0.0 $\vert$ 0.0 $\vert$ 88.9 $\vert$ 100.0\\ 
     
     \hline\hline
     \multicolumn{5}{p{7.5in}}{{\footnotesize \emph{Notes}: The table shows the average percent reduction in daily peak load, average percent increase in daily base load, average reduction in daily standard deviation of load when all ($\alpha=1$), half ($\alpha=0.5$), or a quarter ($\alpha =0.25$) of temperature-sensitive load in each day is shiftable to another hour in the same day using the model with in increased 2$^\circ$C temperature. Baseline load values used were daily averages. These calculations are performed for the individual regions, when regions are pooled within each interconnect, and when all regions across the continental United States are pooled (Nationwide). The calculations are also broken out for Winter and Summer months.}}
    \end{tabular}
 \end{sidewaystable}

\renewcommand{\thetable}{S7}
\begin{table}[h]
    \centering 
    \caption{\textbf{Changes in overall peak and base load, and overall variability, when demand is flattened within each day for different levels of $\alpha$ and regional aggregation, after accounting for demand shift from 2$^\circ$C warming. Related to Figure 4.}}
     \label{tab:smoothdem_stat_overall_2C}
   \begin{tabular}{l c c c}
     \hline\hline
     Level of & Peak Load & Base Load & SD\\
     Connectivity & Reduction & Increase & Reduction\\ [0.5ex] 
     \hline 
     {$\alpha$ value} & 0 $\vert$ 0.25 $\vert$ 0.5 $\vert$ 1 & 0 $\vert$ 0.25 $\vert$ 0.5 $\vert$ 1 & 0 $\vert$ 0.25 $\vert$ 0.5 $\vert$ 1 \\
     \hline
     \multicolumn{4}{c}{-All Seasons-}\\
     \hline
     Regional BA & 0.0 $\vert$ 7.8 $\vert$ 14.5 $\vert$ 15.0 & 0.0 $\vert$ 17.5 $\vert$ 23.1 $\vert$ 28.6 & 0.0 $\vert$ 10.7 $\vert$ 16.5 $\vert$ 18.0\\ 
     Interconnect & 5.4 $\vert$ 13.1 $\vert$ 20.4 $\vert$ 20.4 & 5.9 $\vert$ 11.3 $\vert$ 15.6 $\vert$ 20.0 & 8.5 $\vert$ 18.5 $\vert$ 23.5 $\vert$ 24.1\\ 
     Nationwide & 10.0 $\vert$ 16.5 $\vert$ 21.8 $\vert$ 21.8 & 10.8 $\vert$ 20.6 $\vert$ 25.1 $\vert$ 26.3 & 14.5 $\vert$ 23.2 $\vert$ 26.8 $\vert$ 27.0\\  
     \hline
     \multicolumn{4}{c}{-Winter—Dec/Jan/Feb-}\\
     \hline
     Regional BA & 0.0 $\vert$ 9.8 $\vert$ 13.0 $\vert$ 14.4 & 0.0 $\vert$ 8.9 $\vert$ 13.2 $\vert$ 16.8 & 0.0 $\vert$ 18.8 $\vert$ 25.8 $\vert$ 29.1\\ 
     Interconnect & 4.3 $\vert$ 12.7 $\vert$ 14.8 $\vert$ 16.3 & 2.6 $\vert$ 12.8 $\vert$ 18.0 $\vert$ 20.8 & 9.5 $\vert$ 30.4 $\vert$ 37.1 $\vert$ 39.8\\ 
     Nationwide & 8.8 $\vert$ 16.0 $\vert$ 16.1 $\vert$ 16.1 & 3.9 $\vert$ 15.4 $\vert$ 20.8 $\vert$ 23.4 & 21.7 $\vert$ 39.9 $\vert$ 42.5 $\vert$ 42.7\\ 
     \hline
     \multicolumn{4}{c}{-Summer—Jun/Jul/Aug-}\\
     \hline
     Regional BA & 0.0 $\vert$ 5.2 $\vert$ 13.5 $\vert$ 14.4 & 0.0 $\vert$ 22.6 $\vert$ 30.4 $\vert$ 35.6 & 0.0 $\vert$ 33.9 $\vert$ 45.2 $\vert$ 46.8\\ 
     Interconnect & 3.8 $\vert$ 14.6 $\vert$ 21.8 $\vert$ 21.8 & 3.6 $\vert$ 15.2 $\vert$ 22.8 $\vert$ 24.6 & 4.4 $\vert$ 44.5 $\vert$ 54.3 $\vert$ 54.6\\ 
     Nationwide & 8.2 $\vert$ 18.3 $\vert$ 23.5 $\vert$ 23.5 & 9.6 $\vert$ 18.6 $\vert$ 25.8 $\vert$ 27.8 & 11.9 $\vert$ 51.8 $\vert$ 59.0 $\vert$ 59.0\\  
     \hline\hline
     \multicolumn{4}{p{6in}}{{\footnotesize \emph{Notes}: The table shows the average percent reduction in overall peak load, average percent increase in overall base load, average reduction in overall standard deviation of load when all ($\alpha=1$), half ($\alpha=0.5$), or a quarter ($\alpha =0.25$) of temperature-sensitive load in each day is shiftable to another hour in the same day using the model with in increased 2$^\circ$C temperature. Baseline load values used were overall averages. These calculations are performed for the individual regions, when regions are pooled within each interconnect, and when all regions across the continental United States are pooled (Nationwide). The calculations are also broken out for Winter and Summer months.}}
    \end{tabular}
 \end{table}

\end{document}